\documentclass[useAMS,usenatbib]{mn2e}

\usepackage{graphicx}
\usepackage{amsmath}
\usepackage{bm}
\usepackage{epsfig}
\usepackage{amssymb}
\usepackage{natbib}
\usepackage{threeparttable} 
\usepackage{subfigure}
\usepackage{setspace}
\def\simmore{\mathbin{\lower 3pt\hbox
     {$\rlap{\raise 5pt\hbox{$\char'076$}}\mathchar"7218$}}}   

\title[Iron line and continuum of 4U 1636$-$53]{Iron-line and continuum variations in the XMM-Newton and Suzaku spectra of the neutron-star low-mass X-ray binary 4U 1636$-$53}

\author[Ming Lyu et al.]
{Ming Lyu$^1$\thanks{E-mail: m.lyu@astro.rug.nl}, Mariano M\'endez$^1$, Andrea Sanna$^2$, Jeroen Homan$^3$,
\newauthor
Tomaso Belloni$^4$ and Beike Hiemstra$^1$ \\
$^1$Kapteyn Astronomical Institute, University of Groningen, PO BOX 800, NL-9700 AV Groningen, the Netherlands\\
$^2$Dipartimento di Fisica, Universit\`a degli Studi di Cagliari, SP Monserrato-Sestu km 0.7, I-09042 Monserrato, Italy\\
$^3$MIT Kavli Institute for Astrophysics and Space Research, 70 Vassar Street, Cambridge, MA 02139, USA \\
$^4$INAF - Osservatorio Astronomico di Brera, Via E. Bianchi 46, I-23807 Merate, Italy}

\begin{document}

\date{Accepted XXXX. Received XXXX; in original form XXXX}

\maketitle

\label{firstpage}

\begin{abstract}
We used six simultaneous {\em XMM-Newton} and {\em Rossi X-ray Timing Explorer} plus five {\em Suzaku} observations to study the continuum spectrum and the iron emission line in the neutron-star low-mass X-ray binary 4U 1636$-$53. We modelled the spectra with two thermal components (representing the accretion disc and boundary layer), a Comptonised component (representing a hot corona), and either a Gaussian or a relativistic line component to model an iron emission line at $\sim 6.5$ keV. For the relativistic line component we used either the {\sc diskline}, {\sc laor} or {\sc kyrline} model, the latter for three different values of the spin parameter. The fitting results for the continuum are consistent with the standard truncated disc scenario. We also find that the flux and equivalent width of the iron line first increase and then decrease as the flux of the Comptonised component increases. This could be explained either by changes in the ionisation state of the accretion disc where the line is produced by reflection, or by light bending of the emission from the Comptonised component if the height at which this component is produced changes with mass accretion rate.

\end{abstract}


\section{Introduction}

Low-mass X-ray binaries (LMXBs) consist of a compact object (a neutron star or a black hole) and a late-type companion star with a mass of less than $\sim $1 M$_{\odot}$. Material from the outer layers of the companion is stripped off and accretes onto the compact object via the inner Lagrangian point and an accretion disc \citep{accretion_book}. The inner parts of the accretion disc, and in the case of neutron stars the neutron-star surface and boundary layer, emit mostly in the X-ray band. These systems show also high-energy (up to a few 100 keV) emission, likely produced by Comptonisation of the soft X-ray photons in a hot electron corona \citep{mcc2000,sanna13}. An extra hard tail has been observed in some LMXBs, e.g., GX 17$+$2, 4U 1636$-$53, GX 349$+$2, Sco X-1, 4U 1608$-$522, GX 13$+$1, 4U 1705$-$44 \citep{salvo00,fiocchi06,paizis06,piraino07}.

Based on X-ray spectral and rapid variability properties, \citet{hasinger89} classified the neutron-star low-mass X-ray binaries into two categories: the Z sources and the Atoll sources, owing their names to the shapes that the source traces in an X-ray colour-colour diagram. The Z sources have higher luminosities ($\sim 0.5-1~L_{\rm Edd}$) than the Atoll sources \citep[$0.01-0.2 ~L_{\rm Edd}$; e.g.,][]{done07,homan07,ford2000}. In Atoll sources the X-ray spectrum softens and the time-scale of the majority of the variability components decreases as the luminosity of the source generally increases \citep{hasinger89,van03}. More specifically, at low luminosities these sources are in the so-called hard state, in which the Comptonised component dominates the energy spectrum; this component can be reasonably described by a power law with a photon index of $\sim 1.6 - 2.5$ \citep{yoshida93,mendez97}. The temperature at the inner edge of the accretion disc is relatively low, $0.3 - 0.5$ keV \citep{sanna13}, and the disc usually contributes less than 20\% of the emission in the $1-20$ keV energy range. In the truncated disc scenario \citep[see, e.g.,][and references therein]{done07}, the accretion disc is truncated at a large radius in the hard state, that being the reason for the relatively low inner-disc temperature and thermal component flux \citep{gierlinski03}. At high luminosities, in the soft state, the disc emission in the $1-20$ keV range becomes more significant. The standard accretion-disc model \citep{ss73} predicts that in this case the disc extends down to the innermost stable circular orbit radius, leading to a high disc temperature and a strong thermal component. Compared to the hard state, in the soft state the inner-disc temperature increases to $0.7-1.0$ keV \citep{sanna13}; since the number of soft photons increases as $T^4$, the electrons in the corona are efficiently cooled down via the inverse Compton scattering process, and the Comptonised spectrum steepens \citep[photon index $\sim 2 - 2.5$, e.g.,][]{miya93,mendez97}.
In the soft state, the thermal components dominate the X-ray spectrum below $\sim 20$ keV, and little hard emission is detected (Gierli\'nski \& Done. 2003). Both in the hard and the soft state the neutron star surface or boundary layer, usually fitted with a blackbody component \citep{white88,sanna13}, contributes significantly to the emission at energies below $\sim 10$ keV. Going from the hard to the soft state, or vice-versa, the spectra of these sources display some intermediate properties, and the source is said to be in the intermediate, or transitional, state. (These basic states receive several names depending on the class of the source; see, e.g., \citealt{hasinger89}). The mechanism driving the transition between the hard to the soft state is still unclear. However, it is generally assumed that the change of mass accretion rate and the disc geometry are connected to the state evolution \citep{hasinger89,mendez99,lin07}. 
                
Besides the emission components described above, the accretion disc is likely illuminated by the Comptonised photons and the thermal spectrum from the neutron star and its boundary layer, and as a consequence it produces a reflection spectrum \citep[e.g.][]{fabian10}. Due to the high abundance and fluorescence yield, an iron emission line at $\sim$6$-$7 keV may appear in the spectrum, with an intrinsic line width of the order of 1eV \citep{basko78}. The iron line profile is asymmetrically broadened by the fast disc rotation (Doppler-broadening) and special and general relativistic effects (e.g., Doppler boosting and gravitational redshift) near the central compact object \citep{fabian89}. The final line profile is determined by parameters of the system, like the inclination angle of the disc with respect to the line of sight, the inner radius of the disc, and the spin parameter of the central object. Thus measurements of the iron line profile provide an excellent way to study the physics and geometry of the accretion process \citep{bhatta07,cackett08,miller13}.  

4U 1636$-$53 is an Atoll LMXBs, consisting of a neutron star and a 0.4 M$_{\odot}$ companion in a 3.8 hr orbit \citep{pedersen82}, at 6 kpc distance \citep{galloway06}. The source shows the full range of spectral states \citep{belloni07,alta08}. Highly coherent burst oscillations indicate that the system harbors a millisecond pulsar with a spin frequency of 581 Hz \citep{zhang97,stro02}. A pair of quasi-periodic oscillations (QPOs) at kHz frequencies were discovered by \citet{zhang96} and \citet{wijnands97}. \citet{kaaret99} and \citet{marcio13} found that the soft X-ray emission in the lower kHz QPO (from the pair of kHz QPOs the one at lower frequency) lags the hard X-ray emission, suggesting that the emission is due to reprocessing of hard X-rays in a cooler Comptonising corona or the accretion disc, in a region with a size of at most a few kilometres. In the last ten years 4U 1636--53 showed a regular state transition cycle of $\sim$40 days \citep{shih05,belloni07}, making it an excellent source to study the variations of the broadband spectrum and iron line as a function of spectral state.

In this work, we study the spectrum of 4U 1636$-$53 in different states. We use eleven observations from Suzaku, XMM-Newton, and the Rossi X-Ray Timing Explorer (RXTE) satellites covering a wide range in luminosity, allowing us to investigate the evolution of the different spectral components as a function of spectral state, and study possible correlations between the continuum spectrum and the iron line. Since the disc is illuminated and photon-ionised by the continuum emission to produce the iron line, correlations between the iron line flux and the flux of different continuum components may provide an important clue to understand the origin of the iron emission line and the evolution of the accretion flow geometry.

\section{Observations and data reduction}    

The data used in this work were collected by three satellites: five observations were made with Suzaku, while the other six were made simultaneously with the XMM-Newton and RXTE satellites. We call the Suzaku observations S1 to S5 and the XMM-Newton/RXTE observations X1 to X6, respectively. We provide details of all observations in Table \ref{overall}.

\begin{figure}
\centering
\includegraphics[width=90mm]{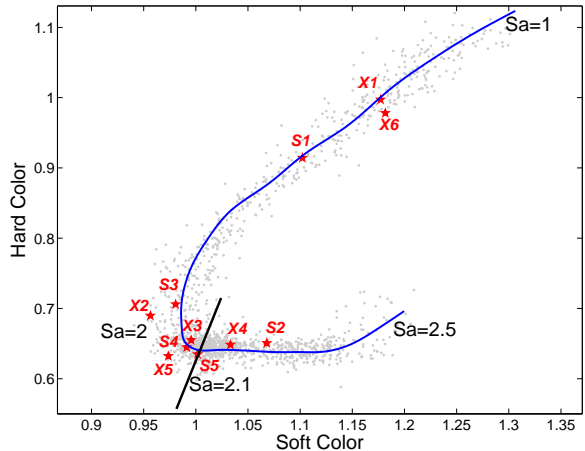}
\caption{Colour-colour diagram of 4U 1636$-$53 using all RXTE observations (see text). Each gray point represents the averaged
 Crab-normalised colours \citep[see][for details]{zhang11} of a single RXTE observation. The red stars 
 mark the position of the five Suzaku observations (S1-S5) and six XMM-Newton/RXTE observations 
 (X1-X6). The position of the source in the diagram is parameterised by the length of the blue solid 
 curve $S_{a}$. The diagonal black solid line indicates the position in the diagram for which
 $S_a=2.1$ \citep[see also][]{zhang11}.}
\label{ccd}
\end{figure}

\begin{figure*}
\centering
\includegraphics[width=1.0\textwidth]{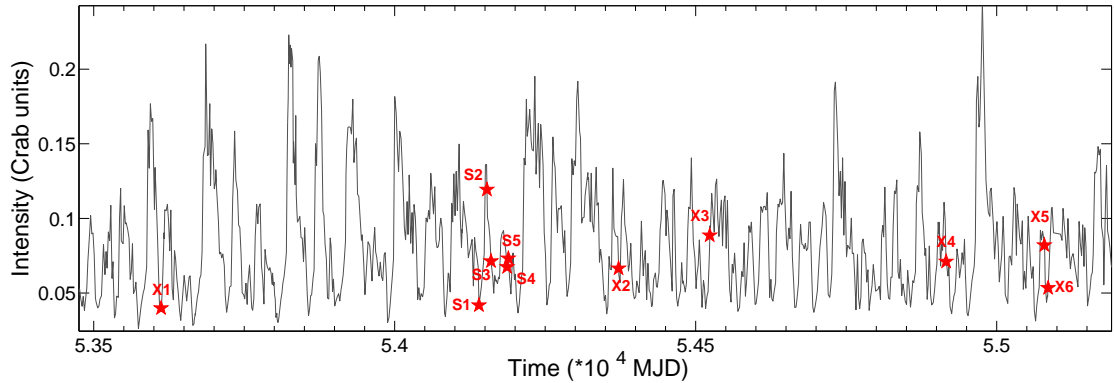}
\caption{Long-term light curve of 4U 1636$-$53 obtained with the RXTE/PCA. For clarity, we only show the part of the light curve around the dates of the XMM-Newton and Suzaku observations. As in Figure \ref{ccd}, the red stars mark the position of the five Suzaku observations (S1-S5) and six XMM-Newton/RXTE observations (X1-X6). 
 }
\label{intensity}
\end{figure*}

We used all 16-s time-resolution Standard-2 data available from the RXTE Proportional Counter Array (PCA) from 1996 February 28 to 2011 September 18 to calculate two X-ray colours and the X-ray light curve. We defined the hard and soft colours as the $9.7-16.0/6.0-9.7$ keV and the $3.5-6.0/2.0-3.5$ keV count-rate ratios, and the intensity as the $2-16$ keV count rate. We normalised the colours and intensity to those of the Crab nebula in observations taken close in time to the ones of 4U 1636--53 \citep[see][for details of the procedure]{zhang09}. We used the parameter $S_{a}$ to indicate the approximate location of the source in the colour-colour diagram \citep[see e.g.][]{mendez_99,zhang11}. We defined $S_{a}$ as in \citet{zhang11}, such that $S_{a}=1$ corresponds to the top right vertex and $S_{a}=2$ to the bottom left vertex of the colour-colour diagram. The coordinate of $S_{a}$ is usually assumed to be a function of mass accretion rate \citep{hasinger89,zhang11}. The $S_{a}$ values of the eleven observations run from 1.33 to 2.23, with three observations (X1, X6 and S1) in the hard branch, $S_{a} < 2 $, two observations (X4 and S2) in the soft branch, $S_{a} > 2$, and the rest clustered around the vertex between the two branches, $S_{a}\sim2$. In Figure \ref{ccd} we show the colour-colour diagram, calculated as described above, for all RXTE observations of 4U 1636--53. We also plot there the curve that we used to define the $S_{a}$ value of each observation. We also indicate there the approximate position of the source in the colour-colour diagram during the eleven observations analysed in this paper. In Figure \ref{intensity} we show part of the light curve of 4U 1636--53 obtained with the PCA detector on board RXTE. We show the eleven observations discussed in this paper following the same convention as in Figure \ref{ccd}.

\subsection{Suzaku data reduction}

The Suzaku observations of 4U 1636$-$53 were carried out using two detectors: the X-ray Imaging Spectrometer (XIS) and the Hard X-ray Detector (HXD). The three XIS detectors cover the $0.2 - 12$ keV energy range, with the two front-illuminated (FI) CCD detectors (XIS0, XIS3) being sensitive in the $0.4-12$ keV range, while the back-illuminated (BI) CCD detector (XIS1) covers the $0.2-12$ keV range. The HXD-PIN camera provides spectra in the $10-70$ keV range. The 2$\times$2 and 3$\times$3 editing modes were applied to the XIS detectors with the 1/4 window mode and a burst option to limit the photon pile-up. For all five observations the XIS-pointing position was applied. We list other details of the observations in Table \ref{overall}.

All data reduction proceeded according to the Suzaku Data Reduction Guide\footnote{http://heasarc.nasa.gov/docs/suzaku/analysis/abc/}. We calibrated the XIS and HXD data using XIS CALDB 20111109 and HXD CALDB 20070710, respectively. We put together the $2\times2$ and $3\times3$ mode event files for each XIS detector into the HEASOFT tool {\tt xselect} to produce good time intervals excluding X-ray bursts, and extracted spectra from a rectangular box region centred at the position of the source. We used the HEASOFT tools {\tt xisrmfgen} and {\tt xissimarfgen} to create the XIS redistribution matrix files (RMFs) and ancillary response files (ARFs). We produced the final FI spectrum file by combining the spectra and responses of the XIS0 and the XIS3 data, while the final BI spectrum file was created from spectra and responses of the XIS1 data. We rebinned all the spectra to a minimum of 20 counts per bin. To test for pile-up, we extracted several spectra from the event files excluding a small rectangular area of different sizes within the rectangular region that we used previously to extract the spectrum of the source. We determined the optimal size of the inner rectangular area by checking that the model parameters of the spectral continuum did not change significantly when we increased the size of the inner rectangular area further. For the FI (BI) spectrum of observation S1 the width and the height of the inner rectangular area were, respectively, 43 and 54 (41 and 39) pixels. For the other four observations we used the same width and height of the inner rectangular area for the FI and BI spectra, respectively, 77 and 103 pixels for S2 and S3, 88 and 88 pixels for S4, and 58 and 82 pixels for S5. 

To produce the HXD-PIN spectra we followed exactly the steps recommended by the Suzaku team. For each observation the HXD-PIN data reduction process began with a clean event file. We extracted the PIN spectrum and applied the dead-time correction using the pseudo-events files. After we had extracted the non-X-ray background (NXB) spectrum, we increased the exposure time of the background spectrum by a factor 10 since the NXB event file was calculated with a count rate 10 times higher than the real background count rate to reduce the statistical errors. Since the cosmic X-ray background was not included in the NXB event file, it was simulated and modelled (as a power law with a high energy cutoff component; see the Suzaku Data Reduction guide for details), and finally it was added to the NXB spectrum in order to provide the total background spectrum. As in the case of the XIS spectra, we excluded the time intervals with X-ray bursts to produce the X-ray spectra.

\subsection{XMM-Newton and RXTE data reduction}

In this paper we used the same XMM-Newton and RXTE data as presented in \citet{sanna13}. Details of the extraction and reduction process are given in \citet{sanna13}. Here we only summarise their steps. The six XMM-Newton observations used in this work were taken with the EPIC-PN cameras in timing mode, in which one dimension of the CCDs is compressed to obtain a fast read out. The PN event files were processed using the tool {\tt epproc} in SAS version 12.1. Intervals with X-ray bursts were excluded from the analysis, and all source spectra were rebinned to ensure there were at least 25 counts in every bin.

For the RXTE observations taken simultaneously with the XMM-Newton observations, \citet{sanna13} extracted spectra from the Standard 2 data taken with the third Proportional Counter Unit (PCU2), since it is the best-calibrated detector, excluding intervals with X-ray bursts. A 0.6$\%$ systematic error was added to the PCA data and the background spectrum was estimated using the tool {\tt pcabackest}. The HEXTE spectrum was produced after excluding X-ray bursts, and no systematic errors were applied. For more details of the XMM-Newton and RXTE data reduction, we refer to \citet{sanna13}.

\begin{table*}
\small
\caption{Suzaku and XMM-Newton/RXTE observations of 4U 1636$-$53.}
\begin{tabular}{|c|c|c|c|c|c|}
\hline
\hline
Observation   &    Instrument    &     ObsID    &       Start Date    &    Start Time     &    Exposure (ks)$^{*}$       \\
\hline
   S1         &     Suzaku       &    401050010 &       09-02-2007    &    09:18:22	  &  22.3(FI); 11.1(BI); 20.3(HXD)                      \\
\\
   S2         &     Suzaku       &    401050020 &  	22-02-2007    &    07:05:38       &  39.7(FI); 19.8(BI); 33.2(HXD)                      \\
\\
   S3         &     Suzaku       & 401050030    &       01-03-2007    &    01:01:07       &  38.2(FI); 19.1(BI); 45.2(HXD)                      \\
\\
   S4         &     Suzaku       & 401050040    &       27-03-2007    &    11:31:36       &   30.3(FI); 14.9(BI); 26.1(HXD)                     \\
\\
   S5         &     Suzaku       &  401050050   &       29-03-2007    &    11:19:47       &  9.8(FI); 4.9(BI); 11.0(HXD)                     \\
\hline
   X1         &    XMM-Newton    &  0303250201  &   29-08-2005        &  18:24:23         &    25.7               \\
              &    RXTE          & 91027-01-01-000 &  &                16:35:28  &26.2 (PCA); 9.0 (HEXTE)        \\
\\
X2         &    XMM-Newton    & 0500350301   &   28-09-2007        &  15:44:56       &        14.3            \\
              &    RXTE          & 93091-01-01-000 & & 14:47:28 & 26.9 (PCA); 8.8 (HEXTE)                 \\
\\
X3         &    XMM-Newton    & 0500350401   &   27-02-2008        &   04:15:37        &     34.7              \\
              &    RXTE          & 93091-01-02-000 & & 03:46:56 &  25.3 (PCA); 8.3 (HEXTE)       \\
\\
X4         &    XMM-Newton    & 0606070201   &    25-03-2009       &   22:59:30        &      23.8              \\
              &    RXTE          & 94310-01-02-03 &  & 23:00:32 & 1.9 (PCA)   \\
              &    RXTE          & 94310-01-02-04 & 26-03-2009 & 00:39:28 & 1.6 (PCA)            \\
              &    RXTE          & 94310-01-02-05 & & 02:17:36 & 1.4 (PCA)    \\
              &    RXTE          & 94310-01-02-02 & & 03:54:24 & 1.3 (PCA); 2.2 (HEXTE)    \\
\\
X5         &    XMM-Newton    & 0606070301   &    05-09-2009  &  01:57:03       &    32.8                \\
              &    RXTE          & 94310-01-03-000 & & 01:17:36 & 16.6 (PCA)     \\
              &    RXTE          & 94310-01-03-00 & & 08:20:32 & 7.3 (PCA); 7.6 (HEXTE)    \\
\\
X6         &    XMM-Newton    & 0606070401   &   11-09-2009  &   08:48:11  &   21.1       \\
              &    RXTE          & 94310-01-04-00 & & 08:42:24 & 18.4 (PCA); 5.7 (HEXTE)     \\
   \hline

\end{tabular}
\medskip
\\
$\ast$ The final exposure time excludes X-ray bursts, background flares and instrument dropouts. The XMM-Newton/RXTE data are the same as those used by \citet{sanna13}.
\label{overall}
\end{table*}

\section{Spectral Analysis}

For each of the Suzaku observation, we fitted the XIS ($0.9-10$ keV) and HXD ($10-30$ keV) spectra together using XSPEC v12.8.0c. We ignored the energy range $1.5-2.5$ keV due to problems in the calibration of the instrument \citep{rivers10,ushio09,garson10}. For the HXD spectrum we selected the energy range below 30 keV since, for 4U 1636--53, that is the energy range where the source emission dominates over the background. The fits of the XMM-Newton/RXTE spectra were done using XSPEC v12.7.1 from $0.8-120$  keV \citep[PN: $0.8-11$ keV; PCA: $10-25$ keV; HEXTE: $20-120$ keV; see][]{sanna13}.

For the Suzaku observations, we used the component {\sc phabs} to describe the photoelectric absorption in the interstellar medium, using the abundance table of \citet{ander89} and the cross section table of \citet{bcmc}. In order to account for possible systematic errors in the cross-calibration between the different instruments, we included a constant factor in all the models. We fixed this constant factor to 1 for the FI and PN spectra in the Suzaku and XMM-Newton data, respectively, and we left it free for all the other spectra. For details of the XMM-Newton and RXTE spectral analysis, we refer to \citet{sanna13}.

\subsection{Spectral model}

\subsubsection{Continuum spectrum}

We needed to include two thermal components and a Compotonised component to obtain a good fit to the continuum spectrum of all observations. We used a multi-colour disc blackbody component, {\sc diskbb} \citep{mitsuda84,maki86}, in XSPEC to describe the emission from the accretion disc, and a single temperature blackbody component, {\sc bbody}, to model the emission from the neutron star surface and boundary layer. For the hard component we used a thermally Comptonised continuum model, {\sc nthcomp} \citep{zdzi96,zyck99}. In this model the source of the thermal seed photons for the Compton process can either come from the disc blackbody or the blackbody component. After some tests, we selected the former as the source of soft photons, the reason being that when we chose the blackbody as the source of seed photons, the fitting results showed that the blackbody component was no longer needed to fit the data \citep[see also][for a similar situation regarding the fits to the XMM-Newton/RXTE data]{sanna13}. 

Previous works have shown that the parameters of the spectral components vary across the different spectral states: The temperatures of the thermal components is typically in the range of $~0.5-2.0$ keV in the soft state \citep{ooster01,disalvo2000a,iaria05}, and below $\sim$ 1 keV in the hard state \citep{church01,barret03}. The temperature of the corona changes in the opposite way, from $2-3$ keV in the soft state to a few tens of keV in the hard state \citep[e.g.,][]{gierlinski03}.  

\subsubsection{Iron line model}

The fits using only a continuum model revealed residuals around $6-7$ keV, the energy range in which an iron emission line could in principle be present, therefore we also included in our models a component in that energy range to account for this putative line. We used, separately, four different models to fit the line (see Table \ref{sline} and \ref{xline}). Firstly, we used a simple {\sc gauss} model to extract some general properties of the emission line, like the significance and the width. Since the resulting Gaussian line was in general broad ($\sigma$ between $\sim 0.7$ keV and $\sim 1.4$ keV), we then also fitted the emission line with two relativistically broadened line models, {\sc diskline} \citep{fabian89} and {\sc laor} \citep{laor91}, developed to describe the emission from a line emitted from an accretion disc around a non-rotating and maximally-rotating black hole, respectively. Because of this last restriction, these two models may not be readily applicable to a moderately fast spinning neutron star. We therefore also used the {\sc kyrline} model \citep{dovciak04} to fit the line, in which the line profile is calculated for the space-time around a black hole with arbitrary spin. In all cases we constrained the energy of the line to the range $6.4-6.97$ keV, corresponding to the energy of the Ly-$\alpha$ line of neutral and H-like iron, respectively. For the {\sc diskline}, {\sc laor} and {\sc kyrline} models we fixed the outer disc radius to $1000 R_{\rm g}$, where $R_{\rm g}=GM/c^{2}$, with $G$ being Newton's constant, $c$ the speed of light, and $M$ the mass of the neutron star. For the {\sc kyrline} model, we did three separate fits with the spin parameter fixed at 0, 0.27 and 1, respectively \citep[see][for details]{sanna13}.

\section{Results}

\subsection{Fitting results}

\begin{figure}
\centering
\includegraphics[height=0.52\textwidth]{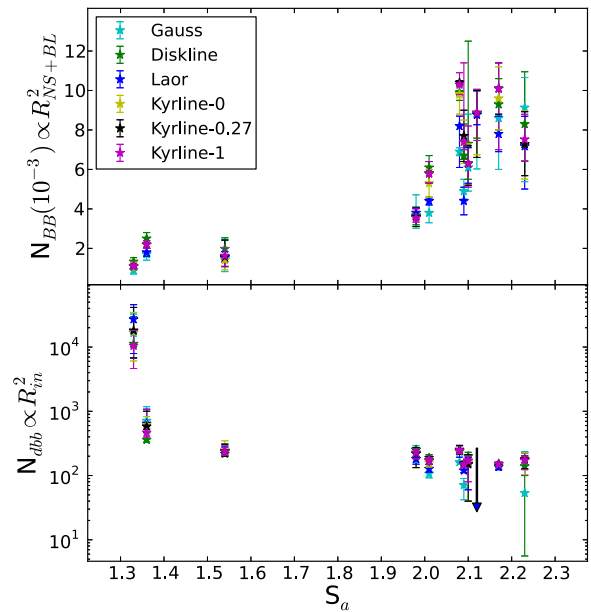}
\caption{Normalisation of the blackbody (upper panel) and disc blackbody (lower panel) components in 4U 1636$-$53 as a function of S$_{a}$ for the five Suzaku and six XMM-Newton/RXTE observations. Different colours show the results of fits with different models to the iron line, as indicated in the legend, with Kyrline-0, Kyrline-0.27 and Kyrline-1 representing the results when we fitted the line with a {\sc kyrline} model with the spin parameter fixed to 0, 0.27 or 1, respectively. The blue arrow in the lower panel shows the upper limit of the normalization in observation S5 where we did not significantly detect emission from the disc.}
\label{Noralization}
\end{figure}

\begin{figure}
\centering
\includegraphics[height=0.58\textwidth]{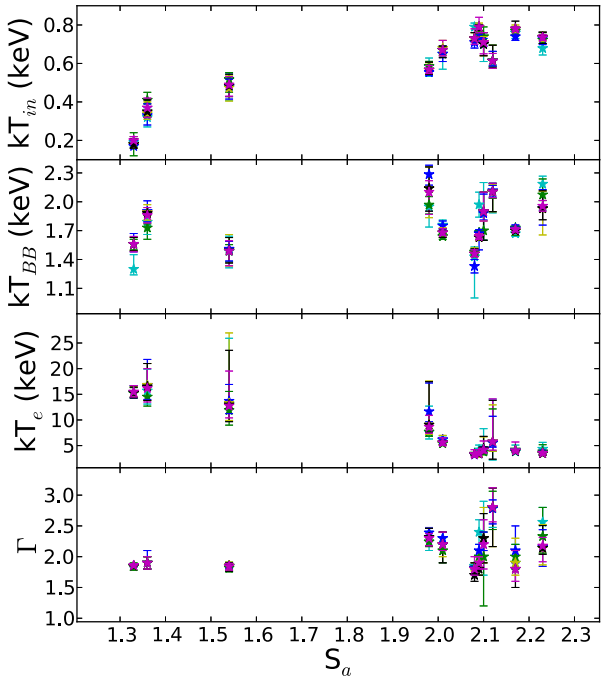}
\caption{Temperature of the {\sc diskbb} (upper panel) {\sc bbody} (second panel from the top) components, electron temperature (third panel from the top) and power-law index $\Gamma$ (lower panel) of the {\sc nthcomp} component in 4U 1636$-$53 as a function of $S_{a}$. Symbols are the same as in Figure \ref{Noralization}.}
\label{Temperature}
\end{figure}

\begin{figure}
\centering
\includegraphics[width=78mm]{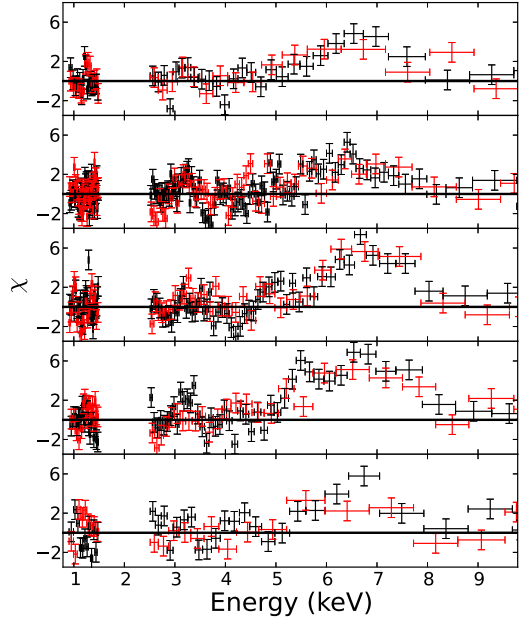}
\caption{Residuals in terms of sigmas for the 5 Suzaku observations of 4U 1636$-$53 for the continuum model described in the text plus a Gaussian line, after we set the normalisation of the {\sc gauss} component in the best fitting model to zero. From top to bottom the plots correspond to the observation S1-S5. The residuals of the FI and BI spectrum are rebinned and plotted in black and red, respectively.}
\label{profile}
\end{figure}

\begin{figure}
\centering
\includegraphics[height=0.42\textwidth]{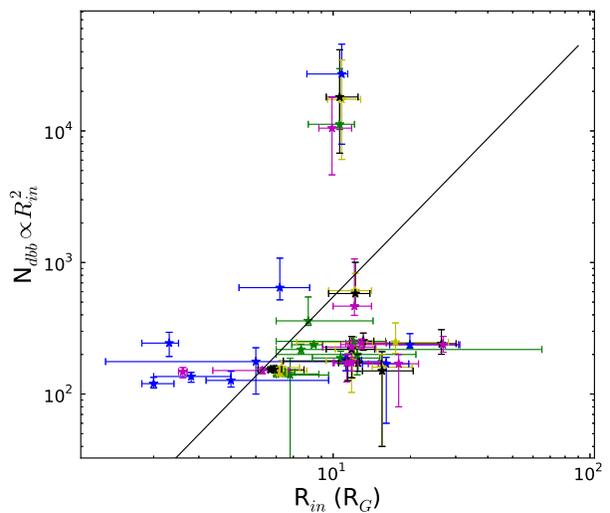}
\caption{Normalisation of the {\sc diksbb} component vs. the inner disc radius deduced from the iron line component for all observations of 4U 1636--53. The black solid line shows the best fitting result when a quadratic relation $N_{dbb} = C \times R_{in}^2$ is fitted to the data. We do not include observation S5 since for that observation the normalization of {\sc diskbb} is consistent with zero. Symbols are the same as in Figure \ref{Noralization}.}
\label{Rin}
\end{figure}

We give the best-fitting parameters to the Suzaku spectra of 4U 1636$-$53 in Tables \ref{scon} and \ref{sline}. For completeness, we also reproduce here the results of \citet{sanna13} for the XMM-Newton and RXTE data (see Tables \ref{xcon} and \ref{xline}; unless otherwise indicated, all errors and upper limits correspond to the 90\% and 95\% confidence intervals for one parameter, respectively). We plot the evolution of the normalisation of the blackbody component, $N_{\rm BB}$, and that of the disc blackbody, $N_{\rm DBB}$, in Figure \ref{Noralization}. The normalisation of the blackbody component is proportional to $(R_{\rm NS + BL})^2$, where $R_{\rm NS + BL}$ is the apparent radius of the neutron star plus boundary layer \citep{gilf03,white88}. The increase of the normalisation of the {\sc bbody} component suggests that, as the source moves from the hard to the soft state, either the true emitting area on the surface of the neutron star increases or the colour-correction factor, which accounts for hardening of the spectrum arising from electron scattering in the neutron-star atmosphere and boundary layer \citep{london86,madej04},
decreases, or both. The {\sc diskbb} normalisation is $N_{\rm DBB} \propto R_{\rm in}^2 \cos{i}/d^2$, where $R_{\rm in}$ is the radius of the inner edge of the accretion disc, $i$ is the inclination of the disc with respect to the line of sight, and $d$ is the distance to the source. Since $d$ is constant, and assuming that the inclination angle of the accretion disc is the same in all observations, $\sqrt{N_{\rm DBB}} \propto R_{\rm in}$. Figure 3 shows that the inferred inner radius of the disc decreases by a factor of $\sim 10$ when the source evolves from the hard to the soft state, and in the soft state $R_{\rm in}$ remains more or less constant, suggesting that the disc may have reached the innermost stable circular orbit (ISCO). The normalisation of the {\sc nthcomp} component (not plotted) shows no significant changes as a function of the source state.

In the soft state, the temperature of the inner disc is $\sim 0.6-0.8$ keV, that of the neutron star surface plus boundary layer is $\sim 1.7-2.0$ keV, and the temperature of the corona is $\sim 2-8$ keV. In the hard state the temperature of the inner disc decreases to less than $\sim 0.5$ keV, the temperature of the neutron-star surface plus boundary layer is $\sim 1.2 - 1.9$ keV, while the temperature of the corona increases to $\sim 15 - 20$ keV. 

Combining the Suzaku results with those from the XMM-Newton/RXTE observations of \citet{sanna13}, in Figure \ref{Temperature} we show the evolution of these three temperatures and the photon index of the {\sc nthcomp} component as a function of $S_{a}$. The temperature at the inner disc radius increases significantly from $\sim 0.2$ keV in the hard state to about 0.8 keV in the soft state (Figure \ref{Temperature}, upper panel), while the temperature of the blackbody increases slightly from $\sim 1.5$ to $\sim 2$ keV when $S_{a}$ increases (Figure \ref{Temperature}, second panel from the top). The inner disc temperature evolution is consistent with the standard truncated disc scenario in neutron star LMXBs \citep{done07}: When the source evolves from the hard to the soft state, mass accretion increases, the inner edge of the accretion disc moves toward the central compact object, and since the temperature in the disc is $T\propto \dot M^{1/4} R^{-3/4}$ \citep{ss73}, the temperature at the inner edge of the disc increases. The increase of the (colour) temperature of the blackbody component could be due to either heating up of the surface of the neutron star plus boundary layer as $\dot M$ increases, or to a decreasing colour-correction factor, as the mass accretion rate onto the neutron star increases. 

Moving from the hard to the soft state, the temperature of the corona (Figure \ref{Temperature}, third panel from the top) drops from $\sim 15$ keV to less than $\sim 5$ keV while $\Gamma$, the power-law index\footnote{The power-law photon index of {\sc nthcomp}, $\Gamma$, and the optical depth, $\tau$, and electron temperature, $kT_e$, of the Compton corona are related through the equation $\Gamma = \sqrt{\frac{9}{4} + \frac{m_e c^2}{kT_e\tau(1+\tau/3)}} - \frac{1}{2}$ \citep{sunyaev80,zdzi96}, where $m_e$ is the electron mass and $c$ the speed of light.} of the {\sc nthcomp} component (Figure \ref{Temperature}, bottom panel) remains more or less constant at around 2, or increases slightly, consistent with past measurement in other sources \citep[e.g.][]{mendez97}. In the soft state the thermal component dominates the emission below $\sim 20$ keV, while in the hard state it is the Comptonised component that dominates the spectrum.

In Figure \ref{profile}, we show the residuals (in terms of sigmas) of all five Suzaku spectra when we fitted them with the continuum model described above plus a Gaussian line in the $6-7$ keV range, with the normalisation of the Gaussian line subsequently set to zero. We examined the significance of the line using the F-test for an extra parameter for the {\sc gauss} model of the line \citep[see][for the applicability of the F-test in this case]{prota02}. The F-test probability for observations S1 to S5 is $0.0094$, $0.0029$, $1.1\times10^{-7}$,  $0.0003$ and $0.0086$, respectively. 

As we described in the previous section, since the Gaussian used to fit the line was always very broad, we also fitted the line with three different relativistically broadened components. (We note, however, that the fits with a Gaussian are good, and the fits with the relativistic line models are not significantly better than those with a Gaussian.) We found that, regardless of the model that we used to fit the line ({\sc diskline}, {\sc laor}, or {\sc kyrline}, the latter with the spin parameter fixed to either 0, 0.27 or 1), in most of the Suzaku observations the inclination angle and the emissivity index of the disc could not be well constrained. Considering this, we give the best-fitting parameters of the line for the different models in Table \ref{sline} (with the exception of the emissivity index, which was completely unconstrained in almost all fits), and we only describe the evolution of the inner disc radius and the flux of the line obtained from the fits. We notice that the best-fitting inclination angle in most cases is larger than $\sim 75 - 80$ degrees. This value is inconsistent with the fact that the source shows no dipping or eclipses, but it is similar to the inclination angle obtained by \citet{pandel08} and \citet{sanna13} from fits to the XMM-Newton spectra of 4U 1636--53.


In Figure \ref{Rin} we plot the normalisation of the {\sc diksbb} component vs. the inner disc radius deduced from the iron line component for all observations of 4U 1636--53 and for the different models used to fit the iron line. The solid line in this plot is the best-fitting quadratic relation to the data (all models of the line together), $N_{dbb} = C \times R_{in}^2$. From this Figure it is apparent that the normalisation of the {\sc diskbb} component and the inner disc radius deduced from the iron line profile do not follow the expected relation.

\subsection{Flux evolution}

\begin{figure}
\centering
\includegraphics[height=0.35\textwidth]{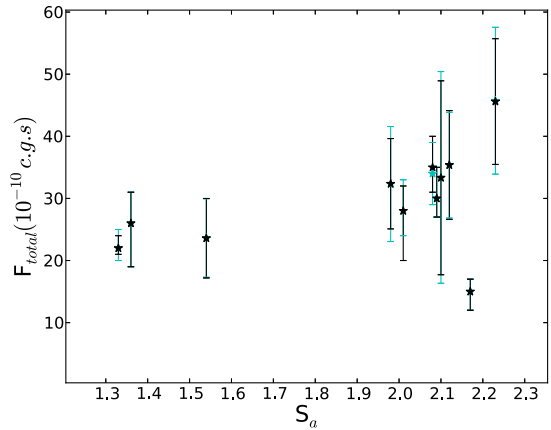}
\caption{Total unabsorbed flux in the $0.5-130$ keV range of 4U 1636$-$53 as a function of $S_{a}$. Symbols are the same as in Figure \ref{Noralization}. For clarity, we only plot the {\sc gauss} and {\sc kyrline-0.27} results here. The other models of the line give consistent results.}
\label{totalflx}
\end{figure}

\begin{figure}
\centering
\includegraphics[height=0.62\textwidth]{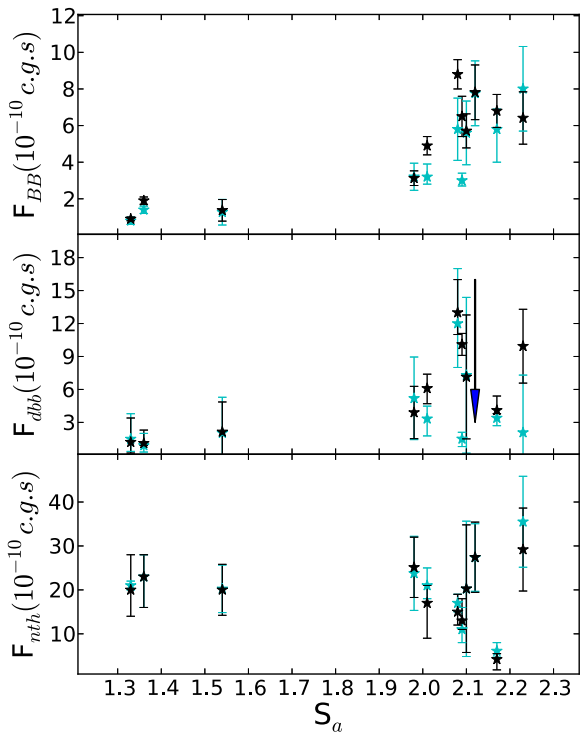}
\caption{Unabsorbed flux for the blackbody and disc blackbody component (upper and middle panels, respectively) in the $0.5-130$ keV range for 4U 1636$-$53 as a function of $S_{a}$. The blue arrow in the second panel shows the upper limit of the flux in observation S5 where we did not significantly detect emission from the disc. The lower panel shows the unabsorbed $0.5-130$ keV flux of the {\sc nthcomp} component in 4U 1636--53 as a function of $S_{a}$. Symbols are the same as in Figure \ref{Noralization}. For clarity, we only plot the {\sc gauss} and {\sc kyrline-0.27} results here. The other models of the line give consistent results.}
\label{conflx}
\end{figure}

\begin{figure}
\centering
\includegraphics[height=0.38\textwidth]{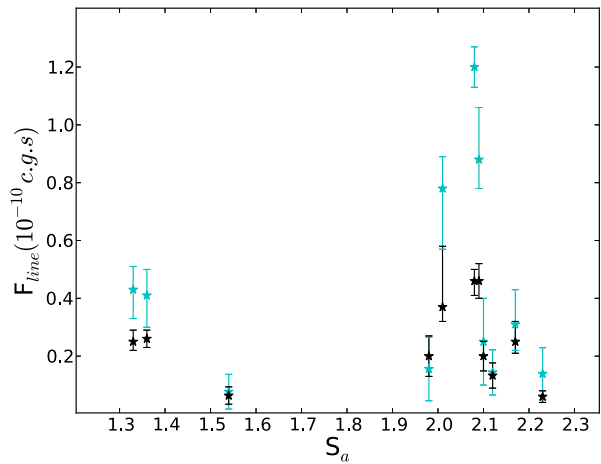}
\caption{Unabsorbed iron line flux in the $0.5-130$ keV range for 4U 1636$-$53 vs. $S_{a}$. Symbols are the same as in Figure \ref{Noralization}. For clarity, we only plot the {\sc gauss} and {\sc kyrline-0.27} results here. The other models of the line give consistent results.}
\label{lineflx}
\end{figure}

As shown in Figure \ref{totalflx}, the $0.5-130$ keV unabsorbed flux (continuum+line) of the eleven observations remains more or less constant, or increases slightly,  as a function of $S_{a}$. From Figure \ref{conflx}, it is apparent that the flux of the blackbody (upper panel) and the disc components (middle panel) both increase with $S_{a}$. The {\sc nthcomp} flux (lower panel) remains more or less constant across the whole $S_{a}$ range. As shown in Figure \ref{lineflx}, the flux of the iron line behaves in a complex way as a function of the $S_{a}$. The line flux initially decreases slightly as $S_{a}$ increases, and it then shows a large scatter at around $S_{a}$=2.1, close to the position of the vertex in the colour-colour diagram (see Figure \ref{ccd}). 

\begin{figure}
\centering
\includegraphics[height=0.75\textwidth]{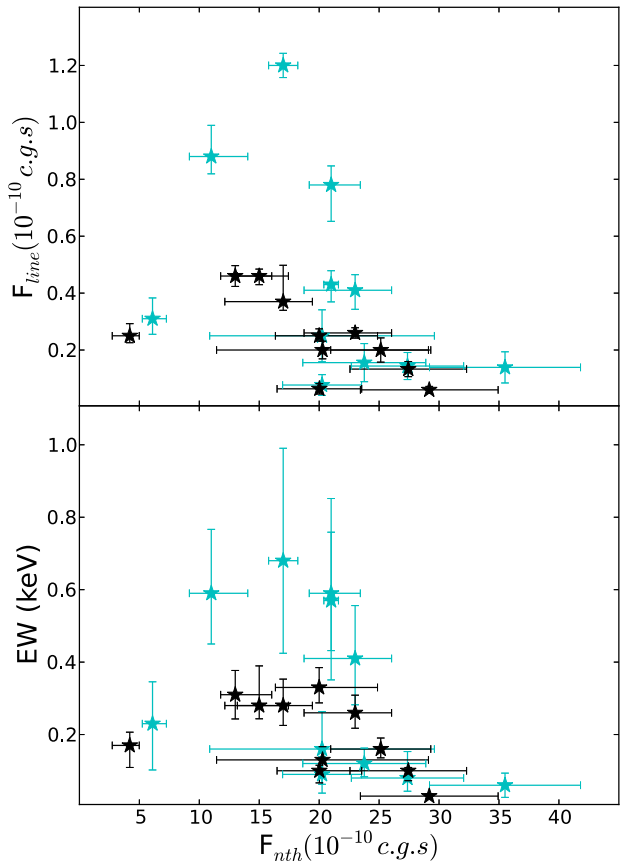}
\caption{Flux ($0.5-130$ keV; upper panel) and equivalent width (lower panel) of the iron line vs. the flux ($0.5-130$ keV) of the {\sc nthcomp} continuum component in 4U 1636$-$53. For clarity, we only plot the results of {\sc gauss} (in cyan) and {\sc kyrline-0.27} (in black) here. The other models of the line give consistent results. For this Figure we use the 1-$\sigma$ errors in the flux and equivalent width.}
\label{corrflx}
\end{figure}

We also studied the correlation between the flux and the equivalent width of the iron line and the two possible illuminating components, either the Comptonising corona (Figure \ref{corrflx}) or the neutron-star surface plus boundary layer. Both panels of Figure \ref{corrflx} show that, for the fits of the line with the {\sc gauss} model, there is a strong positive correlation between, respectively, the flux and equivalent width of the line and the flux of {\sc nthcomp} when the {\sc nthcomp} flux is below $\sim15\times10^{-10}$ erg cm$^{-2}$ s$^{-1}$, and an anti-correlation above that value. For the fit of the line with the the {\sc kyrline} model there is a weak positive correlation between the flux of the line and the flux of the {\sc nthcomp} component when the {\sc nthcomp} flux is below $\sim15\times10^{-10}$ erg cm$^{-2}$ s$^{-1}$, and a strong anti-correlation above that value. For clarity reasons, in both panels of this Figure we only plot the results of the line fits with the {\sc gauss} and {\sc kyrline}, spin parameter $0.27$, models. For both panels, the fits with the {\sc laor} model give results consistent with those of the {\sc gauss} model, while the results for the fits with {\sc diskline} and {\sc kyrline} with spins of 0 and 1 are consistent with those of the {\sc kyrline} with spin 0.27 shown in the Figure. The maximum in this plot happens when the source is at $S_{a} \approx 2.1$, close to the vertex in the colour-colour diagram. The flux and equivalent width of the line do not appear to be correlated with the flux of the {\sc bbody} component (not plotted). Given that the ionising photons that produce the iron line must have energies above $\sim 7$ keV, we also compared the flux and equivalent width of the iron line vs. the $7 - 130$ keV flux of the  {\sc bbody} and {\sc nthcomp} components. The results (not plotted) are similar to those for the full energy range: The flux and equivalent width of the line first increase and then decrease as the $7 - 130$ keV flux of the  {\sc nthcomp} component increases, but are not correlated with the $7 - 130$ keV flux of the  {\sc bbody} component.

\section{Discussion} 

We used all available Suzaku and XMM-Newton observations, the latter complemented with simultaneous RXTE observations, to study the evolution of the continuum spectrum and the iron line emission in the neutron star low-mass X-ray binary 4U 1636$-$53 across different spectral states. We found that the temperature of the neutron star and that at the inner edge of the accretion disc increase, whereas the electron temperature of the corona decreases, as the source moves across the colour-colour diagram from the hard to the soft state. Simultaneously, the inner radius of the accretion disc deduced from the {\sc diskbb} component decreases rapidly by a factor of $\sim 10$, and then remains more or less constant as the source reaches the soft state, suggesting that the inner edge of the disc may have reached the ISCO. The power-law photon index of the component used to fit the corona emission increases from $\sim 1.7$ in the hard state up to $\sim 2.8$ (depending on the model used to fit the line) in the soft state. Simultaneously, the total unabsorbed flux and the flux of the Comptonised component remain more or less constant, whereas the flux of the disc and that of the neutron star plus boundary layer increase, as the source moves from across the colour-colour diagram from the hard to the soft state (all fluxes calculated in the $0.5 -130$ keV range). Interestingly, the flux and equivalent width of the iron line first increase and then decrease as the flux of the Comptonised component increases. The maximum of this relation takes place when the source is close to the vertex in the colour-colour diagram. 

The relation between the flux or the equivalent width of the line on one side and the flux of the hard spectral component on the other appears to contradict the expected behaviour if the line is due to reflection off the accretion disc. In the reflection scenario, the corona (and in neutron-star systems possibly also the neutron-star or boundary-layer) illuminates and photon-ionises the disc, where the reflection spectrum (continuum + emission line) is then produced. In this scenario the flux or equivalent width of the iron line should be positively correlated with that of the illuminating component. We find that, while this is the case when at low {\sc nthcomp} flux values, the opposite is true when the {\sc nthcomp} flux is high (see Figure \ref{corrflx}). However, the disc becomes more ionised when the illuminating flux increases, provided that there are enough photons with energies above the ionisation potential of H- and He-like iron. As the flux of the illuminating component increases further, the material in the disc would eventually become fully ionised, and the line will disappear. This scenario agrees with recent calculations of the reflection spectrum of an ionised slab at different ionisation levels  \citep{garcia13}. \cite{garcia13} found (see their Figure 5) that the flux of the iron emission line in the reflected spectrum decreases gradually as the material in the slab is ionised further. Similarly, \cite{matt93} showed that the equivalent width of the iron line in the reflected spectrum of an ionised slab initially increases and then decreases as the flux of the ionising source increases. Our results are generally consistent with this idea \citep[compare our Figure \ref{corrflx} with  Figures 1 and 2 in][]{matt93}. However, in this scenario, as the ionisation state of the disc increases, the fitted energy of the emission line should gradually increase, as the fraction of H-like iron ions relative to the He-like iron ions in the disc increases. Our results do not show any correlation between the fitted energy of the line and the flux of the ionising source (see Tables \ref{sline} and \ref{xline}), regardless of the model we used to fit the line. While this lack of correlation could be due to limitations of the models that we used to fit the iron line, or to the fact that we only fitted the iron line instead of the full reflection spectrum, using the same XMM-Newton data presented here \citet{sanna13} found no correlation between the ionisation parameter of the full reflection model they fitted to the spectra and the position of the source in the colour-colour diagram, reinforcing this conclusion.

Alternatively, the relation between iron line flux or equivalent width and the flux of the hard spectral component in 4U 1636--53 shown in Figure \ref{corrflx} could be interpreted in terms of light bending \citep{miniutti04}. In this model, the observed flux of the hard illuminating source (in our fits the {\sc nthcomp}) depends strongly upon the height of the source of photons in the corona above the disc; the observed flux of the direct emission can change by up to a factor $\sim20$ as the height of the illuminating source changes, even if the luminosity of the illuminating source remains constant. On the other hand, the reflected spectrum of the disc, and hence the flux or equivalent width of the iron line, is much less sensitive to the height of the illuminating source. \cite{miniutti04} find that there are three regimes in which, as the flux of the direct emission increases, the reflection component (and hence the iron line flux or equivalent width) is first correlated, then insensitive, and finally anti-correlated with the flux of the direct emission. \cite{rossi05} found that, in the black-hole candidate XTE J1650--500, as the total flux of the source above 7 keV increases, the flux of the iron line initially remains constant and eventually decreases. They found that the source flux level at which the line flux starts to decrease coincides with the transition from the hard to the soft state in this source. \cite{park04} also found a complex relation between the iron-line flux and the flux of the hard and the soft components in the black-hole candidate 4U 1543--475. In this case, however, the relation does not follow any clear trend, and an interpretation in terms of the light bending model is not apparent. Rossi et al. (2005) suggested the direct and reflected components in XTE J1650--500 could be related to the existence of a radio jet in this source \citep{corbel04}, with possible changes of the height of the base of the jet as the mechanism that drives the changes of the direct emission in XTE J1650--500. It is unclear whether this could also be the case in 4U 1636--53, since this source has not been detected in radio, and hence the existence of a strong jet in this system is at best doubtful \citep{thomas79,russell12}. The lack of a strong radio jet in 4U 1636--53 could also be the reason that the flux of the hard (direct) component in this source changes only by a factor $\sim 5$, whereas in XTE J1650--500 the flux of the direct ionising source changes by $\simmore 10$. We finally note that if the relation between the flux or equivalent width of the iron line and the {\sc nthcomp} flux in 4U 1636--53 is due to light bending, this would be the first case of a neutron-star system in which this effect is observed. The light bending model was developed for the case of a rapidly spinning black hole, and therefore it is unclear whether it would also apply for a moderately spinning neutron star.

In the case of 4U 1636--53 the point at which the flux of the iron line switches from correlated to anti-correlated with the flux of the {\sc nthcomp} component coincides with the vertex in the colour-colour-diagram, at $S_{a} \sim 2.1$. The existence of the vertex in this diagram directly indicates a sudden change of the spectral properties of the source which, going from the hard to the soft state, quickly softens there. We note also that the vertex of the colour-colour diagram is the place where the quality factor of the kHz QPOs \citep{barret06,mendez06} and the intrinsic coherence between the variability in the hard and soft energy bands \citep{marcio13} are the highest. Furthermore, \citet{zhang11} studied 298 type-I X-ray bursts in the 4U 1636$-$53 using RXTE observations, and they found that in this area of the colour-colour diagram most photospheric radius expansion bursts and a super-burst are observed. All these properties suggest that, in this source several properties of the accretion flow change significantly in this area of the colour-colour diagram. 

The changes of the properties of the continuum spectrum of 4U 1636--53 as the source moves in the colour-colour diagram are generally consistent with those of other neutron-star LMXBs. For instance, \citet{farinelli11} measured the power-law index and the electron temperature $kT_{\rm e}$ of the Comptonising component in the neutron star LMXBs Sco X-1, GX 349$+$2, X 1658$-$298, 1E 1724$-$3045, GX 17$+$2, Cyg X$-$2, GX 340$+$0, GX 3$+$1, and GS 1826$-$238, and found that the power-law energy index ($\Gamma -1$ in the {\sc nthcomp} component) remains more or less constant at around 1$\pm$0.2. Using RXTE and BeppoSAX observations, \citet{seifina11,seifina12} showed that in 4U 1728$-$34 and GX 3$+$1, the power-law index also remains almost constant as the temperature of the corona changes dramatically. Recently, \citet{titar13} found that in another neutron star binary, 4U 1820$-$30, the power-law index remains almost constant for different source states. The power-law index in 4U 1636$-$53 changes from $\sim 1.7$ in the hard state to $\sim 2.8$ in the soft state (depending on the model used to fit the iron line), which is a somewhat larger than the variations in the sources studied by \citet[][note, however, that the model that they used to fit the Comptonised component is not the same as the one we used here]{farinelli11}. \citet{titar13} concluded that the power-law index quasi-stability is an intrinsic property of neutron star binaries, which is fundamentally different from that of black-hole binary systems. \citet{seifina11} suggested that this stability of the power-law index happens when the energy released in the corona itself is much higher than the one from the disc intercepted by the corona.

The changes of the continuum spectrum are also broadly consistent with the truncated disc scenario in LMXBs \citep[e.g.][and references therein]{done07}: The temperature of the disc increases and the inner radius of the disc decreases as the inferred mass accretion rate increases. The electron temperature of the {\sc nthcomp} component and the temperature of the neutron-star surface are also consistent with this scenario. It is intriguing, however, that the inner radius of the accretion disc deduced from the relativistically broaden iron line does not follow the same trend. The model of the disc that we used to fit the data does not include spectral hardening, and the iron line profile may be further affected by mechanisms other than relativistic broadening. It remains to be seen whether these effect could explain this discrepancy.

\section*{Acknowledgments}

This research has made use of data obtained from the High Energy Astrophysics Science Archive Research Center (HEASARC), provided by NASA's Goddard Space Flight Center. This research made use of NASA's Astrophysics Data System. LM is supported by China Scholarship Council (CSC), grant number 201208440011. TMB acknowledges support from INAF PRIN 2012-6. We are grateful to Jon Miller for comments on a previous version of this manuscript.

\clearpage
\begin{table*}
\caption{Suzaku observations of 4U 1636$-$53 - Parameters of the continuum components. The different columns correspond to the different models used to fit the iron line. The {\sc diskbb} normalisation is $N_{\rm DBB} = (R_{\rm in}/d_{10})^2 \cos{i}$, with $R_{\rm in}$ the inner disc radius in km, $d_{10}$ the distance to the source in units of 10 Kpc, and $i$ the inclination of the accretion disc with respect to the line of sight. The {\sc bbody} normalisation is $N_{\rm BB} = L_{39}/d_{10}^2$, with $L_{39}$ the source luminosity in units of $10^{39}$ erg s$^{-1}$ and $d_{10}$ the same as for the {\sc diskbb} component. The {\sc nthcomp} normalisation is the flux density in photons cm$^{-2}$ s$^{-1}$ keV$^{-1}$ at 1 keV.}
\tiny
\begin{tabular}{|c|c|c|c|c|c|c|c|c|}
\hline
\hline
  S1 (S$_{a}$=1.54) &Model comp & Parameter  & GAUSS & DISKLINE &LAOR & KYRLINE-0 & KYRLINE-0.27& KYRLINE-1   \\
\hline
  & PHABS &$N_{\rm H}$(10$^{22}$cm$^{-2}$)   &$  0.23 \pm 0.02              $ &  $ 0.22 \pm 0.02                $ & $ 0.23 \pm 0.01               $ & $  0.24  \pm 0.02        $ & $ 0.23 \pm 0.02       $  & $  0.23 \pm 0.01      $\\
  &DISKBB &kT$_{\rm in}$(keV)                &$  0.48 \pm 0.07              $ &  $ 0.52_{-   0.07 }^{+   0.04 } $ & $ 0.48 \pm 0.06               $ & $  0.47  \pm 0.07        $ & $ 0.48 \pm 0.06       $  & $  0.49 \pm 0.05      $\\
  &       &N$_{\rm DBB}$(10$^{2}$)           &$  2.4  \pm 0.5               $ &  $ 2.2  \pm 0.2                 $ & $ 2.4  \pm 0.3                $ & $  2.5   \pm 0.7         $ & $ 2.4  \pm 0.4        $  & $  2.4  \pm 0.4       $\\
  &BBODY  &kT$_{\rm BB}$(keV)                &$  1.5  \pm 0.2               $ &  $ 1.5  \pm 0.1                 $ & $ 1.5 \pm 0.1                 $ & $  1.5   \pm 0.2         $ & $ 1.5  \pm 0.1        $  & $  1.5  \pm 0.2       $\\
  &       &N$_{\rm BB}$(10$^{-3}$)           &$  1.5  \pm 0.8               $ &  $ 2.0  \pm 0.6                 $ & $ 1.5 \pm 0.3                 $ & $  1.4   \pm 0.6         $ & $ 1.6  \pm 0.7        $  & $  1.6  \pm 0.5       $\\
  &NTHCOMP&$\Gamma$                          &$  1.85 \pm 0.75              $ &  $ 1.82 \pm 0.06                $ & $ 1.86 \pm 0.05               $ & $  1.86  \pm 0.07        $ & $ 1.85 \pm 0.07       $  & $  1.84 \pm 0.06      $\\
  &       &kT$_{\rm e}$(keV)                 &$ 13.1_{-    3.4}^{+   12.8}  $ &  $ 11.9 \pm 3.3                 $ & $13.7  \pm 2.8                $ & $  13.3_{-3.3}^{+13.7}   $ & $13.0_{-3.3 }^{+10.6 }$  & $  12.8_{-2.4}^{+6.8} $\\
  &       &N$_{\rm NTH}$                     &$  0.15 \pm 0.03              $ &  $ 0.13 \pm 0.02                $ & $ 0.151_{- 0.02 }^{+ 0.005}   $ & $  0.16  \pm 0.04        $ & $ 0.15 \pm 0.04       $  & $  0.15 \pm 0.03      $\\
  &       $\chi^2_\nu (\chi^2/dof)$        &  &   1.04 (655/627) &  1.05 (655/625) & 1.04 (651/625) & 1.04 (652/625) & 1.04 (652/625) & 1.04 (653/625) \\

\hline
  S2 (S$_{a}$=2.23) &Model comp & Parameter  & GAUSS & DISKLINE &LAOR & KYRLINE-0 & KYRLINE-0.27& KYRLINE-1   \\
\hline
  & PHABS &$N_{\rm H}$(10$^{22}$cm$^{-2}$)   & $ 0.309 \pm 0.007    $  & $ 0.305 \pm 0.007    $ & $ 0.305 \pm 0.005  $ & $  0.304 \pm 0.004  $ & $ 0.304 \pm 0.004  $  &  $ 0.304  \pm 0.004 $\\
  &DISKBB &kT$_{\rm in}$(keV)                & $ 0.68  \pm 0.05     $  & $ 0.72  \pm 0.03     $ & $ 0.73  \pm 0.03   $ & $  0.74  \pm 0.03   $ & $ 0.74  \pm 0.03   $  &  $ 0.74   \pm 0.02  $\\
  &       &N$_{\rm DBB}$(10$^{2}$)           & $ 0.5_{- 0.5}^{+1.8} $  & $ 1.4_{-1.3}^{+0.5}  $ & $ 1.8   \pm 0.7    $ & $  1.8   \pm 0.7    $ & $ 1.8   \pm 0.4    $  &  $ 1.7    \pm 0.4   $\\
  &BBODY  &kT$_{\rm BB}$(keV)                & $ 2.19_{-0.3}^{+0.08}$  & $ 2.07_{-0.06}^{+0.2}$ & $ 1.9   \pm 0.2    $ & $  1.9   \pm 0.3    $ & $ 1.9   \pm 0.2    $  &  $ 1.96   \pm 0.06  $\\
  &       &N$_{\rm BB}$(10$^{-3}$)           & $ 9.1_{-3.8}^{+1.5}  $  & $ 8.3   \pm 2.3      $ & $ 7.2   \pm 1.9    $ & $  7.3   \pm 1.8    $ & $ 7.3   \pm 1.6    $  &  $ 7.5    \pm 1.2   $\\
  &NTHCOMP&$\Gamma$                          & $ 2.6   \pm 0.3      $  & $ 2.3   \pm 0.4      $ & $ 2.2   \pm 0.3    $ & $  2.1   \pm 0.4    $ & $ 2.1   \pm 0.3    $  &  $ 2.2    \pm 0.2   $\\
  &       &kT$_{e}$(keV)                     & $ 4.3   \pm 1.2      $  & $ 3.8   \pm 1        $ & $ 3.5   \pm 0.5    $ & $  3.5   \pm 0.4    $ & $ 3.5   \pm 0.4    $  &  $ 3.5    \pm 0.3   $\\
  &       &N$_{\rm NTH}$                     & $ 0.57_{-0.2}^{+0.07}$  & $ 0.42  \pm 0.09     $ & $ 0.36  \pm 0.09   $ & $  0.34  \pm 0.1    $ & $ 0.4   \pm 0.1    $  &  $ 0.36   \pm 0.05  $\\
  &      $\chi^2_\nu (\chi^2/dof)$    && 1.17 (733/628) & 1.17 (730/626) & 1.17 (730/626) & 1.16 (728/626) & 1.16 (728/626) & 1.16 (728/626) \\
\hline
  S3 (S$_{a}$=1.98) &Model comp & Parameter  & GAUSS & DISKLINE &LAOR & KYRLINE-0 & KYRLINE-0.27& KYRLINE-1   \\
\hline
  & PHABS &$N_{\rm H}$(10$^{22}$cm$^{-2}$)   & $ 0.292 \pm 0.009  $  & $  0.292 \pm 0.005  $ & $ 0.296 \pm 0.005     $ & $  0.294 \pm 0.007    $ & $ 0.294 \pm 0.006    $  &  $      0.295 \pm 0.006     $\\
  &DISKBB &kT$_{\rm in}$(keV)                & $ 0.59  \pm 0.05   $  & $  0.59  \pm 0.03   $ & $ 0.55  \pm 0.02      $ & $  0.57  \pm 0.03     $ & $ 0.57  \pm 0.03     $  &  $      0.57  \pm 0.02      $\\
  &       &N$_{\rm DBB}$(10$^{2}$)           & $ 2.5_{-0.9}^{+0.4}$  & $  2.5   \pm 0.2    $ & $ 1.8   \pm 0.3       $ & $  2.2   \pm 0.7      $ & $ 2.2   \pm 0.7      $  &  $      2.3   \pm 0.4       $\\
  &BBODY  &kT$_{\rm BB}$(keV)                & $ 2.0   \pm 0.3    $  & $  2.0   \pm 0.1    $ & $ 2.3   \pm 0.1       $ & $  2.1   \pm 0.3      $ & $ 2.1   \pm 0.2      $  &  $      2.1   \pm 0.2       $\\
  &       &N$_{\rm BB}$(10$^{-3}$)           & $ 3.7_{-0.6}^{+1.1}$  & $  3.7   \pm 0.5    $ & $ 3.8   \pm 0.3       $ & $  3.6   \pm 0.5      $ & $ 3.6   \pm 0.5      $  &  $      3.6   \pm 0.4       $\\
  &NTHCOMP&$\Gamma$                          & $ 2.2   \pm 0.2    $  & $  2.24  \pm 0.07   $ & $ 2.39_{-0.01}^{+0.09}$ & $  2.3   \pm 0.1      $ & $ 2.3   \pm 0.1      $  &  $      2.30  \pm 0.09      $\\
  &       &kT$_{e}$(keV)                     & $ 7.6_{-1.3}^{+5.1}$  & $  7.7   \pm 1.1    $ & $11.6_{-2.0}^{+5.6}   $ & $  9.1_{-1.9}^{+8.5}  $ & $ 9.0_{-1.4}^{+8.4}  $  &  $      8.6_{-1.2}^{+2.3}   $\\
  &       &N$_{\rm NTH}$                     & $ 0.3   \pm 0.1    $  & $  0.31  \pm 0.07   $ & $ 0.39  \pm 0.03      $ & $  0.36  \pm 0.1      $ & $ 0.36  \pm 0.09     $  &  $      0.35  \pm 0.07      $\\
  &        $\chi^2_\nu (\chi^2/dof)$   && 1.24 (780/628) & 1.24 (774/626) & 1.23 (771/626) & 1.23 (773/626) & 1.23 (773/626) & 1.23 (773/626)\\
\hline
  S4 (S$_{a}$=2.10)&Model comp & Parameter  & GAUSS & DISKLINE &LAOR & KYRLINE-0 & KYRLINE-0.27& KYRLINE-1   \\
\hline
  & PHABS &$N_{\rm H}$(10$^{22}$cm$^{-2}$)   & $ 0.292 \pm 0.009  $  & $ 0.289 \pm 0.008     $ &$ 0.292 \pm 0.006     $ & $  0.291 \pm 0.009    $ &$  0.292 \pm 0.008    $  & $ 0.292 \pm 0.007     $\\
  &DISKBB &kT$_{\rm in}$(keV)                & $ 0.70  \pm 0.08   $  & $  0.74 \pm 0.04      $ &$  0.71 \pm 0.02      $ & $   0.71 \pm 0.05     $ &$   0.70 \pm 0.06     $  & $  0.71 \pm 0.06      $\\
  &       &N$_{\rm DBB}$(10$^{2}$)           & $ 1.5_{-1.5}^{+0.6}$  & $   2.0 \pm 0.5       $ &$   1.7_{-1.1}^{+0.2} $ & $    1.6_{-1.2}^{+0.4}$ &$    1.5_{-1.1}^{+0.6}$  & $  1.7_{-0.9}^{+0.3}  $\\
  &BBODY  &kT$_{\rm BB}$(keV)                & $ 1.9   \pm 0.3    $  & $   1.7 \pm 0.2       $ &$  1.88_{-0.07}^{+0.2}$ & $    1.9 \pm 0.2      $ &$    1.9 \pm 0.3      $  & $  1.9  \pm 0.2       $\\
  &       &N$_{\rm BB}$(10$^{-3}$)           & $ 6.1_{-1.2}^{+2.7}$  & $   7.3_{-1.8}^{+5.2 }$ &$   6.3 \pm 0.9       $ & $    6.3 \pm 1.1      $ &$    6.3 \pm 1.0      $  & $  6.3  \pm 1.5       $\\
  &NTHCOMP&$\Gamma$                          & $ 2.3   \pm 0.6    $  & $   2.0 \pm 0.6       $ &$   2.3 \pm 0.2       $ & $    2.3 \pm 0.5      $ &$    2.3 \pm 0.4      $  & $  2.2  \pm 0.4       $\\
  &       &kT$_{e}$(keV)                     & $ 4.4_{-0.9}^{+3.9}$  & $   3.8 \pm 0.8       $ &$   4.2 \pm 0.5       $ & $    4.3_{-0.7}^{+2.4}$ &$    4.4_{-0.8}^{+2.4}$  & $  4.2_{-0.6}^{+1.7}  $\\
  &       &N$_{\rm NTH}$                     & $ 0.3   \pm 0.2    $  & $  0.14_{-0.13}^{+0.2}$ &$ 0.24_{-0.05}^{+0.1} $ & $  0.24_{-0.08}^{+0.2}$ &$    0.3 \pm 0.2      $  & $  0.2  \pm 0.1       $\\
  &       $\chi^2_\nu (\chi^2/dof)$  &&    1.09 (807/740) & 1.10 (810/738) &  1.08 (800/738) & 1.08 (799/738) & 1.08 (799/738) & 1.08 (799/738)  \\
\hline
 S5 (S$_{a}$=2.12) &Model comp & Parameter  & GAUSS & DISKLINE &LAOR & KYRLINE-0 & KYRLINE-0.27& KYRLINE-1   \\
\hline
   & PHABS &$N_{\rm H}$(10$^{22}$cm$^{-2}$)  & $ 0.32 \pm 0.01        $  & $  0.32 \pm 0.01    $ & $  0.320\pm 0.006   $ & $  0.32  \pm 0.01    $  & $   0.32 \pm 0.01    $  & $   0.32 \pm 0.01   $\\
   &DISKBB &kT$_{\rm in}$(keV)               & $ 0.62_{-0.02}^{+0.08} $  & $  0.61 \pm 0.04    $ & $  0.61 \pm 0.03    $ & $  0.61  \pm 0.06    $  & $   0.61 \pm 0.06    $  & $   0.61 \pm 0.04   $\\
   &       &N$_{\rm DBB}$(10$^{2}$)          &   <1.93                   &    <1.56              &    <1.43              &    <2.00                &     < 2.00              &        < 1.64        \\
   &BBODY  &kT$_{\rm BB}$(keV)               & $ 2.11_{-0.2}^{+0.09}  $  & $  2.10 \pm 0.07    $ & $  2.12  \pm 0.06   $ & $ 2.10_{-0.2}^{+0.08}$  & $ 2.10_{-0.2}^{+0.09}$  & $   2.10 \pm 0.08   $\\
   &       &N$_{\rm BB}$(10$^{-3}$)          & $ 8.8_{-2.8}^{+1.2}    $  & $  8.8  \pm 1.2     $ & $  8.8_{-0.5}^{+1.2}$ & $  8.9_{-2.1}^{+1.1} $  & $   8.9_{-2.3}^{+1.1}$  & $   8.9  \pm 1.3    $\\
   &NTHCOMP&$\Gamma$                         & $ 2.8  \pm 0.3         $  & $  2.8  \pm 0.4     $ & $2.78_{-0.25}^{+0.14}$& $  2.8   \pm 0.5     $  & $   2.8  \pm 0.5     $  & $   2.8  \pm 0.3    $\\
   &       &kT$_{e}$(keV)                    & $ 5.7_{-3.5}^{+8.4}    $  & $5.6_{-1.7}^{+6.5}  $ & $  5.5_{-0.8}^{+5.2}$ & $  5.8_{-1.8}^{+7.2} $  & $   5.8_{-3.4}^{+8.1}$  & $  5.8_{-1.7}^{+8.3}$\\
   &       &N$_{\rm NTH}$                    & $ 0.52_{-0.3}^{+0.01}  $  & $0.52_{-0.1}^{+0.02}$ & $0.52_{-0.2}^{+0.01}$ & $ 0.52_{-0.3}^{+0.01}$  & $ 0.52_{-0.3}^{+0.01}$  & $   0.52 \pm 0.01   $\\
   &       $\chi^2_\nu (\chi^2/dof)$  &&  1.04 (635/613) & 1.04 (633/611) & 1.03 (632/611) & 1.03 (632/611) & 1.03 (632/611) & 1.03 (632/611)   \\
\hline

\label{scon}
\end{tabular}
\end{table*}

\clearpage
\begin{table*}
\caption{XMM-Newton/RXTE observations of 4U 1636$-$53 - Parameters of the continuum components. The different columns correspond to the different models used to fit the iron line. The normalisations are the same as in Table \ref{scon}}
\tiny
\begin{tabular}{|c|c|c|c|c|c|c|c|c|}
\hline
\hline
  X1 (S$_{a}$=1.33) &Model comp & Parameter  & GAUSS & DISKLINE &LAOR & KYRLINE-0 & KYRLINE-0.27& KYRLINE-1   \\ 
\hline
  & PHABS &$N_{\rm H}$(10$^{22}$cm$^{-2}$)   &$  0.42 \pm 0.08        $  & $  0.34 \pm 0.07     $ & $   0.4  \pm 0.1     $ &$    0.41_{-0.11}^{+0.06}    $ & $   0.40 \pm 0.09       $  & $   0.47 \pm 0.05    $\\
  &DISKBB &kT$_{\rm in}$(keV)                &$  0.19 \pm 0.03        $  & $  0.18 \pm 0.06     $ & $   0.17 \pm 0.02    $ &$    0.18 \pm 0.03           $ & $   0.18 \pm 0.03       $  & $   0.20 \pm 0.02    $\\
  &       &N$_{\rm DBB}$(10$^{2}$)           &$ 166.5_{-99.9}^{+159.9}$  & $112.4_{-10}^{+185.3}$ & $   271  \pm 189     $ &$    173.9_{- 113.2}^{+171.9}$ & $181.1_{-113.4}^{+232.8}$  & $ 105.0_{-58.6}^{+75.8}$\\
  &BBODY  &kT$_{\rm BB}$(keV)                &$  1.3_{-0.06}^{+0.15}  $  & $  1.56 \pm 0.06     $ & $   1.56 \pm 0.09    $ &$    1.56 \pm 0.07           $ & $   1.56 \pm 0.07       $  & $   1.56 \pm 0.08    $\\
  &       &N$_{\rm BB}$(10$^{-3}$)           &$  0.9  \pm 0.2         $  & $  1.3  \pm 0.2      $ & $   1.1  \pm 0.2     $ &$    1.1  \pm 0.2            $ & $   1.1  \pm 0.1        $  & $   1.1  \pm 0.1     $\\
  &NTHCOMP&$\Gamma$                          &$  1.85 \pm 0.02        $  & $  1.84 \pm 0.06     $ & $   1.85 \pm 0.01    $ &$    1.85 \pm 0.01           $ & $   1.85 \pm 0.01       $  & $   1.86 \pm 0.01    $\\
  &       &kT$_{\rm e}$(keV)                 &$ 15.4  \pm 1.2         $  & $ 15.2  \pm 1.0      $ & $   15.4 \pm 1.2     $ &$    15.4 \pm 1.0            $ & $  15.3  \pm 1.1        $  & $  15.5  \pm 1.1     $\\
  &       &N$_{\rm NTH}$                     &$  0.198\pm 0.008       $  & $  0.193\pm 0.006    $ & $   0.201\pm 0.008   $ &$    0.199\pm 0.006          $ & $   0.198\pm 0.004      $  & $  0.200 \pm  0.004  $\\
  &       $\chi^2_\nu (\chi^2/dof)$    &           &1.08 (306/284) &1.18 (333/282)& 1.09 (307/282)& 1.11 (312/282) &1.11 (312/282) &1.12 (317/282)\\

\hline
  X2 (S$_{a}$=2.01) &Model comp & Parameter  & GAUSS & DISKLINE &LAOR & KYRLINE-0 & KYRLINE-0.27& KYRLINE-1   \\
\hline
  & PHABS &$N_{\rm H}$(10$^{22}$cm$^{-2}$)   & $ 0.30 \pm 0.05         $  & $ 0.32  \pm 0.02     $ & $ 0.31  \pm 0.03    $ & $  0.31  \pm 0.03     $ & $ 0.31 \pm 0.03 $  &  $      0.32 \pm 0.03   $\\
  &DISKBB &kT$_{\rm in}$(keV)                & $ 0.65 \pm 0.06         $  & $ 0.67  \pm 0.02     $ & $ 0.65  \pm 0.04    $ & $  0.68  \pm 0.04     $ & $ 0.67 \pm 0.03 $  &  $      0.67 \pm 0.04   $\\
  &       &N$_{\rm DBB}$(10$^{2}$)           & $ 1.1  \pm 0.3          $  & $ 1.9   \pm 0.2      $ & $ 1.3   \pm 0.2     $ & $  1.6   \pm 0.2      $ & $ 1.7  \pm 0.3  $  &  $      1.7  \pm 0.2    $\\
  &BBODY  &kT$_{\rm BB}$(keV)                & $ 1.76 \pm 0.05         $  & $ 1.64  \pm 0.03     $ & $ 1.75  \pm 0.05    $ & $  1.68  \pm 0.05     $ & $ 1.68 \pm 0.05 $  &  $      1.68 \pm 0.04   $\\
  &       &N$_{\rm BB}$(10$^{-3}$)           & $ 3.8  \pm 0.7          $  & $ 6.1   \pm 0.5      $ & $ 4.4   \pm 0.2     $ & $  5.3   \pm 0.7      $ & $ 5.8  \pm 0.6  $  &  $      5.8  \pm 0.6    $\\
  &NTHCOMP&$\Gamma$                          & $ 2.3  \pm 0.1          $  & $ 2.1   \pm 0.2      $ & $ 2.3   \pm 0.1     $ & $  2.2   \pm 0.2      $ & $ 2.2  \pm 0.3  $  &  $      2.2  \pm 0.2    $\\
  &       &kT$_{e}$(keV)                     & $ 6.1  \pm 0.6          $  & $ 5.4   \pm 0.2      $ & $ 6.2   \pm 0.3     $ & $5.6_{-0.4}^{+ 1.4 }  $ & $ 5.6  \pm 0.5  $  &  $      5.6  \pm 0.7    $\\
  &       &N$_{\rm NTH}$                     & $ 0.29 \pm 0.04         $  & $ 0.19  \pm 0.02     $ & $ 0.27  \pm 0.05    $ & $  0.21  \pm 0.04     $ & $ 0.21 \pm 0.05 $  &  $      0.21 \pm 0.04  $\\
  &  $\chi^2_\nu (\chi^2/dof)$  & & 1.05 (298/284)& 1.07 (304/282) &1.06 (299/282) & 1.07 (301/282)& 1.07 (303/282) & 1.07 (302/282)\\
\hline
  X3 (S$_{a}$=2.08) &Model comp & Parameter  & GAUSS & DISKLINE &LAOR & KYRLINE-0 & KYRLINE-0.27& KYRLINE-1   \\
\hline
  & PHABS &$N_{\rm H}$(10$^{22}$cm$^{-2}$)   & $ 0.30 \pm 0.01       $  & $ 0.36 \pm 0.06   $ & $ 0.36 \pm 0.01       $ & $  0.36 \pm 0.01    $ & $ 0.37 \pm 0.09  $  & $       0.36 \pm 0.03    $\\
  &DISKBB &kT$_{\rm in}$(keV)                & $ 0.79 \pm 0.05       $  & $ 0.73 \pm 0.01   $ & $ 0.71 \pm 0.03       $ & $  0.73 \pm 0.02    $ & $ 0.73 \pm 0.02  $  & $       0.73 \pm 0.04    $\\
  &       &N$_{\rm DBB}$(10$^{2}$)           & $ 1.62_{-0.03}^{+0.7} $  & $ 2.37 \pm 0.04   $ & $ 2.4  \pm 0.5        $ & $  2.45 \pm 0.04    $ & $ 2.5  \pm 0.4   $  & $       2.4  \pm 0.2     $\\
  &BBODY  &kT$_{\rm BB}$(keV)                & $ 1.4  \pm 0.3        $  & $ 1.48 \pm 0.03   $ & $ 1.33 \pm 0.07       $ & $  1.46 \pm 0.05    $ & $ 1.46 \pm 0.04  $  & $       1.46 \pm 0.04    $\\
  &       &N$_{\rm BB}$(10$^{-3}$)           & $ 6.9  \pm 0.2        $  & $ 9.9  \pm 0.3    $ & $ 8.2_{-2.1}^{+0.5}   $ & $  9.8_{-1}^{+0.1}  $ & $10.4  \pm 0.1   $  & $       10.3 \pm 0.6     $\\
  &NTHCOMP&$\Gamma$                          & $ 1.85 \pm 0.04       $  & $ 1.82 \pm 0.03   $ & $ 1.82 \pm 0.02       $ & $1.80_{-0.03}^{+0.1}$ & $ 1.7  \pm 0.2   $  & $       1.8  \pm 0.2     $\\
  &       &kT$_{\rm e}$(keV)                 & $ 3.3  \pm 0.1        $  & $ 3.4  \pm 0.1    $ & $ 3.2  \pm 0.1        $ & $  3.3  \pm 0.1     $ & $ 3.3  \pm 0.1   $  & $       3.3  \pm 0.7     $\\
  &       &N$_{\rm NTH}$                     & $ 0.15 \pm 0.01       $  & $ 0.13 \pm 0.03   $ & $ 0.15 \pm 0.02       $ & $  0.13 \pm 0.01    $ & $ 0.11 \pm 0.01  $  & $       0.11 \pm 0.04    $\\
  & $\chi^2_\nu (\chi^2/dof)$ &&1.09 (306/281)& 1.19 (333/279)& 1.12 (314/279)&1.16 (325/279)&1.17 (326/279)& 1.17 (328/279)\\
\hline
  X4 (S$_{a}$=2.17) &Model comp & Parameter  & GAUSS & DISKLINE &LAOR & KYRLINE-0 & KYRLINE-0.27& KYRLINE-1   \\
\hline
  & PHABS &$N_{\rm H}$(10$^{22}$cm$^{-2}$)   & $ 0.30 \pm 0.01       $  & $ 0.30  \pm 0.01      $ & $ 0.31 \pm 0.02   $ & $  0.30  \pm 0.03       $ & $ 0.30 \pm 0.01   $  &  $      0.30 \pm 0.01    $\\
  &DISKBB &kT$_{in}$(keV)                    & $ 0.76 \pm 0.03       $  & $ 0.77  \pm 0.02      $ & $ 0.74 \pm 0.02   $ & $  0.78  \pm 0.02       $ & $ 0.78 \pm 0.03   $  &  $      0.78 \pm 0.01    $\\
  &       &N$_{\rm DBB}$(10$^{2}$)           & $ 1.5  \pm 0.1        $  & $ 1.42  \pm 0.06      $ & $ 1.4  \pm 0.1    $ & $  1.50_{-0.04}^{+0.08} $ & $ 1.53 \pm 0.07   $  &  $      1.51 \pm 0.08    $\\
  &BBODY  &kT$_{BB}$(keV)                    & $ 1.67 \pm 0.03       $  & $ 1.73_{-0.08}^{+0.02}$ & $ 1.72 \pm 0.04   $ & $  1.70  \pm 0.02       $ & $ 1.70 \pm 0.02   $  &  $      1.71 \pm 0.02    $\\
  &       &N$_{\rm BB}$(10$^{-3}$)           & $ 8.6  \pm 2.1        $  & $ 9.3   \pm 1.3       $ & $ 7.8  \pm 0.9    $ & $  9.6   \pm 1.6        $ & $10.1  \pm 1.3    $  &  $      10.1_{-3.1}^{+1.3}$\\
  &NTHCOMP&$\Gamma$                          & $ 1.9  \pm 0.2        $  & $ 2.0   \pm 0.2       $ & $ 2.1  \pm 0.3    $ & $  1.9   \pm 0.3        $ & $ 1.8  \pm 0.3    $  &  $      1.8  \pm 0.3     $\\
  &       &kT$_{\rm e}$(keV)                 & $ 3.8_{-0.2}^{+1.3}   $  & $ 4.2   \pm 0.2       $ & $ 4.1  \pm 0.2    $ & $  4.0   \pm 0.2        $ & $ 3.9  \pm 0.2    $  &  $      4.0_{-0.2}^{+1.7}$\\
  &       &N$_{\rm NTH}$                     & $ 0.10 \pm 0.02       $  & $ 0.09  \pm 0.01      $ & $ 0.14 \pm 0.03   $ & $  0.07  \pm 0.01       $ & $ 0.06 \pm 0.01   $  &  $      0.06 \pm 0.02    $\\
  & $\chi^2_\nu (\chi^2/dof)$ && 1.08 (304/281)&1.06 (297/279)&1.09 (304/279)&1.06 (295/279)&1.06 (296/279)&1.06 (296/279)\\

\hline
  X5 (S$_{a}$=2.09) &Model comp & Parameter  & GAUSS & DISKLINE &LAOR & KYRLINE-0 & KYRLINE-0.27& KYRLINE-1   \\
\hline
  & PHABS &$N_{\rm H}$(10$^{22}$cm$^{-2}$)   & $ 0.28 \pm 0.02      $  & $ 0.30 \pm 0.02      $ & $ 0.29 \pm 0.03        $ &$   0.29 \pm 0.02     $ & $ 0.29 \pm 0.01   $  &  $      0.29 \pm 0.01       $\\
  &DISKBB &kT$_{\rm in}$(keV)                & $ 0.75 \pm 0.04      $  & $ 0.75 \pm 0.01      $ & $ 0.73 \pm 0.02        $ &$   0.79 \pm  0.02    $ & $ 0.79 \pm 0.02   $  &  $      0.79 \pm 0.04       $\\
  &       &N$_{\rm DBB}$(10$^{2}$)           & $ 0.7  \pm 0.3       $  & $ 1.46 \pm 0.09      $ & $ 1.20 \pm 0.12        $ &$   1.46 \pm 0.09     $ & $ 1.54 \pm 0.09   $  &  $      1.5  \pm 0.2        $\\
  &BBODY  &kT$_{\rm BB}$(keV)                & $ 2.0  \pm 0.1       $  & $ 1.68 \pm 0.02      $ & $ 1.67_{-0.17 }^{+0.04}$ &$   1.63 \pm 0.03     $ & $ 1.63 \pm 0.02   $  &  $      1.64 \pm 0.03       $\\
  &       &N$_{\rm BB}$(10$^{-3}$)           & $ 4.9  \pm 0.5       $  & $ 6.7_{-0.2}^{+1.6}  $ & $ 4.4  \pm 0.7         $ &$   7.4  \pm 1.1      $ & $ 7.7  \pm 1.3    $  &  $      7.4_{-0.4}^{+4}     $\\
  &NTHCOMP&$\Gamma$                          & $ 2.4  \pm 0.2       $  & $ 2.0  \pm 0.1       $ & $ 2.1  \pm 0.1         $ &$   1.9  \pm 0.2      $ & $ 1.8  \pm 0.1    $  &  $      1.9  \pm 0.1        $\\
  &       &kT$_{\rm e}$(keV)                 & $ 4.2  \pm 0.8       $  & $ 3.6_{-0.8}^{+0.1}  $ & $ 3.4_{-0.1}^{+0.8}    $ &$   3.5_{-0.7}^{+0.2} $ & $ 3.4  \pm 0.1    $  &  $      3.5  \pm 0.1        $\\
  &       &N$_{\rm NTH}$                     & $ 0.29_{-0.04}^{+0.1}$  & $ 0.18 \pm 0.03      $ & $ 0.23 \pm 0.02        $ &$   0.12 \pm 0.01     $ & $ 0.11 \pm 0.02   $  &  $      0.12 \pm 0.01       $\\
  & $\chi^2_\nu (\chi^2/dof)$ &&1.12 (304/272)& 1.22 (331/270)&1.15 (310/270)&1.17 (317/270)&1.18 (320/270)&1.21 (327/270)\\

\hline
  X6 (S$_{a}$=1.36) &Model comp & Parameter  & GAUSS & DISKLINE &LAOR & KYRLINE-0 & KYRLINE-0.27& KYRLINE-1   \\
\hline
  & PHABS &$N_{\rm H}$(10$^{22}$cm$^{-2}$)   & $ 0.32 \pm 0.05     $  &$  0.27_{-0.09 }^{+0.03}$ & $ 0.32 \pm 0.07    $ & $  0.31 \pm 0.05    $ & $ 0.31 \pm 0.06    $  &  $      0.29 \pm 0.05         $\\
  &DISKBB &kT$_{\rm in}$(keV)                & $ 0.32 \pm 0.05     $  &$  0.41 \pm 0.03        $ & $ 0.33 \pm 0.06    $ & $  0.34 \pm 0.05    $ & $ 0.35 \pm 0.05    $  &  $      0.37 \pm 0.05         $\\
  &       &N$_{\rm DBB}$(10$^{2}$)           & $ 7.1_{-2.5}^{+4.7} $  &$  3.6_{-0.3}^{+1.9}    $ & $ 6.4_{-1.2}^{+4.4}$ & $  6.1_{-0.5}^{+2.2}$ & $ 5.8_{-1.1}^{+4.2}$  &  $      4.7_{-0.7}^{+6}       $\\
  &BBODY  &kT$_{\rm BB}$(keV)                & $ 1.8  \pm 0.1      $  &$  1.73_{-0.1}^{+0.04}  $ & $ 1.9  \pm 0.1     $ & $  1.89 \pm 0.08    $ & $ 1.88 \pm 0.02    $  &  $      1.86 \pm 0.08         $\\
  &       &N$_{\rm BB}$(10$^{-3}$)           & $ 1.7  \pm 0.3      $  &$  2.5  \pm 0.3         $ & $ 1.8  \pm 0.2     $ & $  2.2  \pm 0.1     $ & $ 2.2  \pm 0.2     $  &  $      2.2  \pm 0.2          $\\
  &NTHCOMP&$\Gamma$                          & $ 1.9  \pm 0.1      $  &$  1.9  \pm 0.1         $ & $ 1.9  \pm 0.2     $ & $  1.9  \pm 0.1     $ & $ 1.9  \pm 0.1     $  &  $      1.9  \pm 0.1          $\\
  &       &kT$_{\rm e}$(keV)                 & $ 15.7_{-2.6}^{+4.2}$  &$ 14.5  \pm 1.8         $ & $16.8_{-2.8}^{+5}  $ & $ 16.8_{-3}^{+4.2}  $ & $16.5_{-2.7}^{+4.5}$  &  $     16.2_{-2.7}^{+3.8}     $\\
  &       &N$_{\rm NTH}$                     & $ 0.23 \pm 0.02     $  &$  0.19 \pm 0.01        $ & $ 0.23 \pm 0.02    $ & $  0.22 \pm 0.02    $ & $ 0.22 \pm 0.02    $  &  $      0.21 \pm 0.01         $\\
  & $\chi^2_\nu (\chi^2/dof)$  &&0.93 (259/278) &1.06 (291/276)&0.94 (260/276)&0.97 (268/276)&0.97 (268/276)&0.98 (270/276)\\
\hline
\label{xcon}
\end{tabular}
\end{table*}

\begin{table*}
\caption{Suzaku observations of 4U 1636$-$53 - Parameters of the Fe line$^{1}$ for the different models used to fit the line. In all cases the line normalisation is given in units of photons cm$^{-2}$ s$^{-1}$.}
\small
\begin{tabular}{|c|c|c|c|c|c|c|}
\hline
\hline
S1 (S$_{a}$=1.54)        &        $E_{\rm line}$(keV)    &     $\sigma$(keV)     &     $i(^\circ)$         &       R$_{\rm in}$(R$_{\rm g}$)         & Normalisation(10$^{-3}$)     &  $\chi^2_\nu (\chi^2/dof)$      \\
\hline
GAUSS      &   6.40 $_{-    0 }^{+  0.26 }$  &  0.8$ \pm 0.4     $ &             -           &                                      &  0.7 $_{ -0.5 }^{+  0.8  }  $ &  1.04 (655/627)    \\
DISKLINE   &   6.40 $_{-    0 }^{+  0.31 }$  &          -          &   44.9 $_{ - 9.7}^{+   45.1}$ &   7.5$_{  - 1.5  }^{+ 57.5}$   &  0.4$_{  -0.2}^{+   0.4    }$ &  1.05 (655/625) \\
LAOR       &   6.49 $_{- 0.09 }^{+  0.10 }$  &          -          &   86.3 $ \pm 0.3            $ &  19.9$_{  - 3.3  }^{+ 11.2}$   &  0.7$_{  -0.2}^{+   0.3    }$ &  1.04 (651/625) \\
KYR-0      &   6.40 $_{-    0 }^{+  0.19 }$  &          -          &   86.2 $_{ -11.8}^{+    2.8}$ &  17.5$ \pm 10.5            $   &  0.7$_{  -0.3}^{+   0.4    }$ &  1.04 (652/625) \\
KYR-0.27   &   6.52 $_{- 0.12 }^{+  0.05 }$  &          -          &   87.6 $_{ -14.9}^{+    1.3}$ &  26.4$_{  -13.2  }^{+  3.7}$   &  0.6$       \pm 0.2         $ &  1.04 (652/625) \\
KYR-1      &   6.54 $_{- 0.14 }^{+  0.20 }$  &          -          &   87.6 $_{ -57.0}^{+    1.3}$ &  26.8$_{  -16.1  }^{+  3.9}$   &  0.6$_{  -0.1}^{+   0.4    }$ &  1.04 (653/625) \\
\hline
S2 (S$_{a}$=2.23)        &        $E_{\rm line}$(keV)    &    $\sigma$(keV)      &      $i(^\circ)$               &       R$_{\rm in}$(R$_{\rm g}$)   &  Normalisation(10$^{-3}$)  &  $\chi^2_\nu (\chi^2/dof)$        \\
\hline
GAUSS      &   6.40 $_{-    0 }^{+  0.08 }$  & 0.8 $\pm 0.3      $ &                   -            &          -                    &  1.3 $_{ -0.8 }^{+  0.9  } $ & 1.17 (733/628)\\
DISKLINE   &   6.97 $_{- 0.35 }^{+     0 }$  &          -          &   23.8 $\pm 5.0              $ &   6.8$_{  - 0.8  }^{+  2.8}$  &  0.9$ \pm 0.5              $ & 1.17 (730/626)\\
LAOR       &   6.50 $\pm 0.06             $  &          -          &   0.01 $_{ - 0.01}^{+   13.2}$ &   5.0$_{  - 3.7  }^{+ 11.1}$  &  0.5$ \pm 0.4              $ & 1.17 (730/626)\\
KYR-0      &   6.40 $_{-    0 }^{+  0.30 }$  &          -          &   28.2 $_{ - 5.3}^{+    1.3} $ &  11.8$_{  - 2.4  }^{+  0.9}$  &  0.6$ \pm 0.2              $ & 1.16 (728/626)\\
KYR-0.27   &   6.42 $_{- 0.02 }^{+  0.36 }$  &          -          &   27.8 $_{ - 3.8}^{+    1.7} $ &  11.4$_{  - 5.0  }^{+  1.3}$  &  0.6$ \pm 0.2              $ & 1.16 (728/626)\\
KYR-1      &   6.40 $_{-    0 }^{+  0.23 }$  &          -          &   28.3 $_{ - 5.2}^{+    1.2} $ &  11.3$ \pm 1.3             $  &  0.6$ \pm 0.2              $ & 1.16 (728/626)\\
\hline
S3 (S$_{a}$=1.98)        &        $E_{\rm line}$(keV)      &  $\sigma$(keV)      &          $i(^\circ)$         &     R$_{\rm in}$(R$_{\rm g}$)     &   Normalisation(10$^{-3}$)  &  $\chi^2_\nu (\chi^2/dof)$         \\
\hline
GAUSS      &   6.61 $_{- 0.21 }^{+  0.17 }$  & 0.8 $\pm 0.2      $ &                    -          &         -                        &  1.4 $_{ -0.7 }^{+  1.4  }  $ & 1.24 (780/628) \\
DISKLINE   &   6.43 $_{- 0.03 }^{+  0.12 }$  &          -          &   79.3 $_{ -20.9}^{+   10.7}$ &  12.0$ \pm 5.0               $   &  1.4$ \pm 0.4               $ & 1.24 (774/626) \\
LAOR       &   6.44 $_{- 0.04 }^{+  0.11 }$  &          -          &   86.4 $\pm 0.2             $ &  11.25$_{  - 0.07  }^{+  1.7}$   &  2.3$ \pm 0.4               $ & 1.23 (771/626)\\
KYR-0      &   6.40 $_{-    0 }^{+  0.13 }$  &          -          &   85.3 $_{ -10.7}^{+    1.9}$ &  11.9$_{  - 1.7  }^{+  2.8}  $   &  1.9$_{  -0.3}^{+   0.6    }$ & 1.23 (773/626) \\
KYR-0.27   &   6.40 $_{-    0 }^{+  0.13 }$  &          -          &   85.1 $_{ - 5.5}^{+    1.8}$ &  11.8$ \pm 2.5               $   &  1.9$ \pm 0.7               $ & 1.23 (773/626) \\
KYR-1      &   6.44 $_{- 0.04 }^{+  0.12 }$  &          -          &   83.9 $_{ - 8.5}^{+    2.5}$ &  11.8$ \pm 2.8               $   &  1.8$ \pm 0.5               $ & 1.23 (773/626) \\
\hline
 S4 (S$_{a}$=2.10)        &        $E_{\rm line}$(keV)    &   $\sigma$(keV)      &          $i(^\circ)$          &   R$_{\rm in}$(R$_{\rm g}$)      & Normalisation(10$^{-3}$)  &  $\chi^2_\nu (\chi^2/dof)$         \\
\hline
GAUSS      &   6.40 $_{-    0 }^{+0.18   }$  & 1.04 $\pm 0.25    $ &             -                  &           -             &    2.3 $\pm 1.4            $ &1.09 (807/740)  \\
DISKLINE   &   6.40 $_{-    0 }^{+0.17   }$  &          -          &   90.0  $_{- 28.4 }^{+0      }$& 12.4  $_{  -6.4}^{+8.6 }$&   1.2 $\pm 0.5            $ &  1.10 (810/738) \\
LAOR       &   6.55 $_{- 0.11 }^{+0.07   }$  &          -          & 86.26   $\pm 0.2              $& 16.1  $\pm 3.4          $&   1.6 $\pm 0.4            $ & 1.08 (800/738) \\
KYR-0      &   6.46 $_{- 0.06 }^{+0.13   }$  &          -          & 86.9    $\pm 1.5              $& 15.5  $_{  -1.3}^{+4.8 }$&   1.8 $\pm 0.3            $ &  1.08 (799/738)\\
KYR-0.27   &   6.47 $_{- 0.07 }^{+0.13   }$  &          -          & 86.8    $\pm 1.5              $& 15.5  $_{  -2.5}^{+5.0 }$&   1.8 $\pm 0.4            $ &  1.08 (799/738)\\
KYR-1      &   6.55 $_{- 0.12 }^{+0.09   }$  &          -          & 86.1    $\pm 1.5              $& 18.0  $\pm 3.9          $&   1.6 $\pm 0.5            $ &  1.08 (799/738) \\
\hline
S5 (S$_{a}$=2.12)        &        $E_{\rm line}$(keV)    &     $\sigma$(keV)     &          $i(^\circ)$            &   R$_{\rm in}$(R$_{\rm g}$)  &Normalisation(10$^{-3}$)  &  $\chi^2_\nu (\chi^2/dof)$         \\
\hline
GAUSS      &   6.40 $_{- 0 }^{+   0.20}$  & 0.7 $\pm 0.3         $ &             -                   &          -                    &  1.4  $\pm 0.8           $ &1.04 (635/613)   \\
DISKLINE   &   6.40 $_{- 0 }^{+   0.23}$  &    -                   &  76.2 $_{   -22.7}^{+    13.8} $& 26.9$_{   -18.2}^{+    12.7 }$&  1.2  $\pm 0.4           $ &1.04 (633/611)\\
LAOR       &   6.40 $_{- 0 }^{+   0.26}$  &    -                   &  86.2 $_{   -13.4}^{+     2.8} $& 32.4$_{   -24.6}^{+    10.8 }$&  1.3  $\pm 0.3           $ &1.03 (632/611)\\
KYR-0      &   6.41 $_{-0.01}^{+  0.04}$  &    -                   &  86.3 $\pm 2.5                 $& 37.2$_{   -15.2}^{+   852.7} $&  1.2  $\pm 0.4           $ &1.03 (632/611)\\
KYR-0.27   &   6.42 $_{-0.02}^{+   0.04}$  &    -                   &  86.3 $\pm 2.5                 $& 37.2$_{   -15.4}^{+   800.1} $&  1.2  $\pm 0.4           $ &1.03 (632/611) \\
KYR-1      &   6.42 $_{-0.02}^{+   0.05}$  &    -                   &  86.3 $\pm 2.5                 $& 37.0$_{   -16.0}^{+   258.7} $&  1.2  $\pm 0.4           $ &1.03 (632/611)\\
\hline
\label{sline}
\end{tabular}

\noindent
$^{1}$Although it was left free during the fits, we do not give the emissivity index in this (and the next Table) because in almost all fits this parameter was completely unconstrained. The errors in all the other parameters were calculated leaving the emissivity index free.
\end{table*}

\clearpage
\begin{table*}
\caption{XMM-newton/RXTE observations of 4U 1636$-$53 - Parameters of the Fe line, for the different models used to fit the line. In all cases the line normalisation is given in units of photons cm$^{-2}$ s$^{-1}$.}
\small
\begin{tabular}{|c|c|c|c|c|c|c|}
\hline
\hline
X1 (S$_{a}$=1.33)        &        $E_{\rm line}$(keV)    & $\sigma$(keV)         &              $i(^\circ)$         &     R$_{\rm in}$(R$_{\rm g}$)      &          Normalisation(10$^{-3}$)  &  $\chi^2_\nu (\chi^2/dof)$        \\
\hline
GAUSS      &   6.40 $_{-  0   }^{+  0.07 }$  & 1.27 $\pm 0.12    $ &        -                          &      -                            &       4.2$\pm 0.9                 $ & 1.08 (306/284)      \\
DISKLINE   &   6.40 $_{-  0   }^{+  0.02 }$  &          -          &   90.0  $_{-  15.9}^{+      0  }$ &   10.6$_{-    2.6 }^{+   1.5  }$  &       2.1$\pm 0.2                 $ & 1.18 (333/282)  \\
LAOR       &   6.44 $_{- 0.04 }^{+  0.08 }$  &          -          &   86.3  $\pm 0.2                $ &   10.8$_{-    2.9 }^{+   0.6  }$  &       2.5$\pm 0.4                 $ & 1.09 (307/282)   \\
KYR-0      &   6.40 $_{-  0   }^{+  0.06 }$  &          -          &   86.1  $\pm 0.8                $ &   10.8$\pm 1.7                 $  &       2.4$\pm 0.3                 $ & 1.11 (312/282)  \\
KYR-0.27   &   6.40 $_{-  0   }^{+  0.06 }$  &          -          &   85.9  $\pm 0.8                $ &   10.6$_{-    1.2 }^{+   1.9  }$  &       2.4$\pm 0.3                 $ & 1.11 (312/282)  \\
KYR-1      &   6.40 $_{-  0   }^{+  0.06 }$  &          -          &   85.1  $\pm 0.9                $ &   9.9 $_{-    1.1 }^{+   1.9  }$  &       2.5$\pm 0.3                 $ & 1.12 (317/282)  \\
\hline
 X2 (S$_{a}$=2.01)        &        $E_{\rm line}$(keV)    & $\sigma$(keV)        &              $i(^\circ)$         &  R$_{\rm in}$(R$_{\rm g}$)         &        Normalisation(10$^{-3}$)  & $\chi^2_\nu (\chi^2/dof)$        \\
\hline
GAUSS      &   6.40 $_{-  0   }^{+  0.07 }$  & 1.3 $\pm 0.1      $ &         -                         &      -                            &      7.6$\pm 1.6                 $  &  1.05 (298/284) \\
DISKLINE   &   6.40 $_{-  0   }^{+  0.07 }$  &          -          &   72.5  $_{-  11.7}^{+     17.5}$ &   10.7$_{-    2.4 }^{+   4.5  }$  &      3.2$\pm 0.5                 $  &  1.07 (304/282)\\
LAOR       &   6.40 $_{-  0   }^{+  0.24 }$  &          -          &   86.8  $_{-  0.4 }^{+      3.2}$ &   4.0 $_{-    0.8 }^{+   5.6  }$  &      6.4$\pm 0.1                 $  &  1.06 (299/282)\\
KYR-0      &   6.4  $_{-  0   }^{+  0.2  }$  &          -          &   88.9  $\pm 1.0                $ &   6.3 $_{-    0.3 }^{+   1.1  }$  &      5.3$\pm 0.8                 $  &  1.07 (301/282)\\
KYR-0.27   &   6.4  $_{-  0   }^{+  0.1  }$  &          -          &   83.9  $_{-  7.9 }^{+      2.1}$ &   12.5$\pm 2.5                 $  &      3.6$_{-   0.6}^{+     2    }$  &  1.07 (303/282)\\
KYR-1      &   6.4  $_{-  0   }^{+  0.1  }$  &          -          &   83.0  $_{-  5.8 }^{+      2.1}$ &   11.8$_{-    1.7 }^{+   3.2  }$  &      3.6$\pm 0.5                 $  &  1.07 (302/282)\\
\hline
X3 (S$_{a}$=2.08)        &        $E_{\rm line}$(keV)    & $\sigma$(keV)         &              $i(^\circ)$         &       R$_{\rm in}$(R$_{\rm g}$)    &        Normalisation(10$^{-3}$)   &  $\chi^2_\nu (\chi^2/dof)$      \\
\hline
GAUSS      &   6.40 $_{-  0   }^{+  0.06 }$  & 1.4$\pm 0.1       $ &    -                              &           -                       &      11.7 $\pm 0.7                $   & 1.09 (306/281)\\
DISKLINE   &   6.40 $_{-  0   }^{+  0.03 }$  &          -          &   90.0  $_{-  15.9}^{+       0 }$ &   8.4 $_{-    1.5 }^{+   0.7  }$  &      4.6 $\pm 0.4                 $   & 1.19 (333/279)\\
LAOR       &   6.40 $_{-  0   }^{+  0.05 }$  &          -          &   87.5  $_{-  0.5 }^{+      2.5}$ &   2.3 $\pm 0.4                 $  &      9.6$_{-   1  }^{+     0.5  } $   & 1.12 (314/279)\\
KYR-0      &   6.67 $\pm 0.07             $  &          -          &   86.2  $\pm 0.7                $ &   13.1$\pm 1.3                 $  &      4.4$\pm 0.4                  $   & 1.16 (325/279) \\
KYR-0.27   &   6.68 $\pm 0.07             $  &          -          &   86.1  $\pm 0.8                $ &   13.1$\pm 1.4                 $  &      4.3$\pm 0.4                  $   & 1.17 (326/279)\\
KYR-1      &   6.7  $\pm 0.1              $  &          -          &   85.6  $_{-  5.8 }^{+      2.1}$ &   12.9$_{-    1.7 }^{+   3.2  }$  &      4.3$\pm 0.5                  $   & 1.17 (328/279)\\
\hline
X4 (S$_{a}$=2.17)        &        $E_{\rm line}$(keV)    & $\sigma$(keV)         &              $i(^\circ)$         &     R$_{\rm in}$(R$_{\rm g}$)      &    Normalisation(10$^{-3}$)    &  $\chi^2_\nu (\chi^2/dof)$      \\
\hline
GAUSS      &   6.86 $\pm 0.11             $  & 1 $\pm 0.2        $ &      -                            &          -                        &     2.9$\pm 1.0                 $    & 1.08 (304/281) \\
DISKLINE   &   6.43 $_{-  0.03}^{+  0.05 }$  &          -          &   73.4  $\pm 5.3                $ &   6.0 $_{-    0   }^{+   2.8  }$  &     2.6$_{-   0.4}^{+     1.0  }$    & 1.06 (297/279)\\
LAOR       &   6.72 $_{-  0.32}^{+  0.15 }$  &          -          &   88.1  $\pm 1.9                $ &   2.8 $_{-    0.8 }^{+   1.2  }$  &     3.5$\pm 0.1                 $    & 1.09 (304/279)\\
KYR-0      &   6.43 $_{-  0.03}^{+  0.1  }$  &          -          &   72.3  $_{-  4.9 }^{+      9.2}$ &   6.0 $_{-    0   }^{+   1.9  }$  &     2.4$\pm 0.6                 $    & 1.06 (295/279)\\
KYR-0.27   &   6.43 $_{-  0.03}^{+  0.07 }$  &          -          &   72.1  $_{-  5.2 }^{+      9.2}$ &   5.7 $_{-    0.6 }^{+    2   }$  &     2.4$\pm 0.5                 $    & 1.06 (296/279)\\
KYR-1      &   6.44 $_{-  0.04}^{+  0.07 }$  &          -          &   74.3  $\pm 6.5                $ &   5.3 $\pm 1.6                 $  &     2.3$\pm 0.5                 $    & 1.06 (296/279)\\
\hline
X5 (S$_{a}$=2.09)        &        $E_{\rm line}$(keV)    & $\sigma$(keV)         &             $i(^\circ)$          &     R$_{\rm in}$(R$_{\rm g}$)      &       Normalisation(10$^{-3}$)   &  $\chi^2_\nu (\chi^2/dof)$      \\
\hline
GAUSS      &   6.41 $_{- 0.01 }^{+  0.09 }$  &1.4$\pm 0.1        $ &          -               &          -                                 &   8.6   $_{ -1    }^{+   1.8    }$& 1.12 (304/272) \\
DISKLINE   &   6.40 $_{-  0   }^{+  0.02 }$  &          -          &   89.9  $_{-  14.5}^{+      0.1}$ &   6.5 $_{-    0.5 }^{+   0.6  }$  &   3.8$\pm 0.5                 $   & 1.22 (331/270)\\
LAOR       &   6.40 $_{-  0   }^{+  0.06 }$  &          -          &   90.0  $_{-  1.7 }^{+      0  }$ &   2.0 $_{-    0.2 }^{+   0.4  }$  &   6.1$\pm 0.7                 $   & 1.15 (310/270)\\
KYR-0      &   6.40 $_{-  0   }^{+  0.04 }$  &          -          &   87    $\pm 1                  $ &   6.2 $_{-    0.2 }^{+   0.4  }$  &   4.6$\pm 0.5                 $   & 1.17 (317/270)\\
KYR-0.27   &   6.40 $_{-  0   }^{+  0.04 }$  &          -          &   86.7  $\pm 1.5                $ &   5.9 $\pm 0.6                 $  &   4.5$\pm 0.6                 $   & 1.18 (320/270)\\
KYR-1      &   6.40 $_{-  0   }^{+  0.03 }$  &          -          &   89.0  $_{-  1.3 }^{+      1  }$ &   2.6 $\pm 0.1                 $  &   5.4$\pm 0.6                 $   & 1.21 (327/270)\\
\hline
X6 (S$_{a}$=1.36)        &        $E_{\rm line}$(keV)    & $\sigma$(keV)         &              $i(^\circ)$         &    R$_{\rm in}$(R$_{\rm g}$)     &      Normalisation(10$^{-3}$)    &  $\chi^2_\nu (\chi^2/dof)$      \\
\hline
GAUSS      &   6.40 $_{-  0   }^{+  0.05 }$  & 1.2$\pm 0.1       $ &      -                   &         -                                &     4   $\pm 1                   $&  0.93 (259/278)  \\
DISKLINE   &   6.40 $_{-  0   }^{+  0.02 }$  &          -          &   90.0  $_{-  21.5}^{+       0 }$ &   8.0 $_{   - 2   }^{+  6.3   }$ &     2.4 $_{-   0.7}^{+     0.3  }$ &  1.06 (291/276)\\
LAOR       &   6.40 $_{-  0   }^{+  0.07 }$  &          -          &   86.4  $\pm 0.2                $ &   6.2 $\pm 1.9                 $ &    3.7 $\pm 0.6                 $ &  0.94 (260/276)\\
KYR-0      &   6.40 $_{-  0   }^{+  0.07 }$  &          -          &   86.8  $\pm 0.8                $ &   12.2$_{  -  2.6 }^{+  1.9   }$ &    2.6 $\pm 0.3                 $ &  0.97 (268/276) \\
KYR-0.27   &   6.40 $_{-  0   }^{+  0.07 }$  &          -          &   86.5  $\pm 0.8                $ &   12.2$_{  -  2.6 }^{+  1.7   }$ &    2.5 $\pm 0.3                 $ &  0.97 (268/276)\\
KYR-1      &   6.40 $_{-  0   }^{+  0.09 }$  &          -          &   86    $\pm 1                  $ &   12.1$\pm 2.1                 $ &    2.5 $\pm 0.3                 $ &  0.98 (270/276)\\
\hline
\label{xline}
\end{tabular}
\end{table*}

\clearpage
\bibliographystyle{mn}
\bibliography{biblio}


\label{lastpage}

\end{document}